\documentclass[11pt, a4paper]{article}

\usepackage{amsmath}
\usepackage{amsfonts}
\usepackage{amssymb}
\usepackage{bbold}
\usepackage{mathrsfs}
\usepackage{latexsym} 
\usepackage{cancel}
\usepackage{bm}
\usepackage{graphicx, rotating}
\usepackage{epstopdf}
\usepackage{epsfig}
\usepackage{latexsym}
\usepackage{color}
\usepackage[dvipsnames]{xcolor}
\usepackage{cite}
\usepackage{slashed}
\usepackage{hyperref}
\usepackage{comment}
\usepackage[utf8]{inputenc}
\usepackage{soul}
\usepackage{multirow}
\usepackage{graphics,subfigure}
\usepackage[dvipsnames]{xcolor}

\hypersetup{colorlinks, citecolor=bluscuro, linkcolor=black, urlcolor=bluscuro}
\definecolor{bluscuro}{rgb}{0.15, 0.2, .85}

\newcommand{\be}{\begin{equation}}
\newcommand{\ee}{\end{equation}}
\newcommand{\bea}{\begin{eqnarray}}
\newcommand{\eea}{\end{eqnarray}}

\newcommand{\Tr}{\text{Tr}}
\newcommand\blfootnote[1]{%
  \begingroup
  \renewcommand\thefootnote{}\footnote{#1}%
  \addtocounter{footnote}{-1}%
  \endgroup
}

\newcommand{\mybeta}{\beta}
\newcommand{\nobeta}{\cancel \beta}
\newcommand{\mtt}{m_{t \bar t}}
\newcommand{\cosCM}{\cos \theta_\text{CM}}
\newcommand{\thCM}{\theta_\text{CM}}

 \def\bea{\begin{eqnarray}}
  \def\eea{\end{eqnarray}}

	\def \beq {\begin{equation}}
	\def \eeq {\end{equation}}
	\def \ba {\begin{array}}
	\def \ea {\end{array}}

\newcommand{\orcid}{\includegraphics{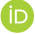}}
\newcommand{\orcidlink}[1]{\href{https://orcid.org/#1}{{\orcid}}}

\begin{document}

\begin{titlepage}
\begin{flushright}
IFT-UAM/CSIC-22-45
\end{flushright}
\begin{center} ~~\\
\vspace{0.5cm} 
\Large { \bf\Large Improved tests of entanglement and Bell inequalities with LHC tops} 
\vspace*{1.5cm}

\normalsize{
{\bf 
J. A. Aguilar-Saavedra\orcidlink{0000-0002-5475-8920}\blfootnote{ja.a.s@csic.es} and
J. A. Casas\orcidlink{0000-0001-5538-1398}\blfootnote{
j.alberto.casas@gmail.com}
 } \\
 
\smallskip  \medskip
\emph{Instituto de F\'\i sica Te\'orica, IFT-UAM/CSIC,}\\
\emph{Universidad Aut\'onoma de Madrid, Cantoblanco, 28049 Madrid, Spain}}

\medskip

\vskip0.6in 

\end{center}

\centerline{ \large\bf Abstract }
\vspace{.5cm}
\noindent
We discuss quantum entanglement in top pair production at the LHC. Near the $t \bar t$ threshold, entanglement observables are enhanced by suppressing the contribution of $q \bar q$ subprocesses, which is achieved by a simple cut on the velocity of the $t \bar t$ system in the laboratory frame. Furthermore, we design new observables that directly measure the relevant combinations of $t\bar t$ spin correlation coefficients involved in the measurement of entanglement and Bell inequalities. As a result, the statistical sensitivity is enhanced, up to a factor of 7 for Bell inequalities near threshold.

\vspace*{2mm}
\end{titlepage}


\section{Introduction}
\label{sec:Intro}

As firstly acknowledged by E. Schr\"odinger, entanglement is {\em the} characteristic aspect of quantum mechanics that enforces its complete departure from classical thought \cite{Schrodinger:1935}. The physical consequence of entanglement that shows such departure is the violation of Bell inequalities \cite{Bell:1964kc} by quantum mechanics, an impossible result for any local and realistic (`classical') theory of nature. In consequence, it is of the utmost importance to experimentally test both issues at different scales. It should be noted here that entanglement does not necessarily imply violation of the Bell inequalities, but the opposite is true.

The violation of Bell inequalities in its Clauser-Horne-Shimony-Holt (CHSH) version \cite{Clauser:1969ny} has been convincingly shown in experiments with low-energy photons and electrons in a number of dramatic experiments \cite{Aspect:1982fx, Hensen:2015ccp}, which have essentially closed all the conceivable loopholes on the validity of the tests. On the other hand such experimental tests have not been possible yet at high-energy scales.

This goal has been recently addressed in several interesting works \cite{Afik:2020onf, Fabbrichesi:2021npl, Severi:2021cnj, Afik:2022kwm,Aoude:2022imd}, dealing with the top-antitop system ($t\bar t$) at the Large Hadron Collider (LHC). As stressed in Ref.\cite{Severi:2021cnj}, the qubits associated to the spin states of $t\bar t$ pairs produced at the LHC provide a suitable arena to investigate these matters since top quarks decay before their spins are randomised by strong radiation and the spin of the lepton produced in semileptonic decays $t \to \ell \nu b$ ($\ell = e,\mu$) is completely correlated to that of the mother top. Besides, owing to the large cross section, $\sigma = 832$ pb at 13 TeV~\cite{Czakon:2011xx} there is a good amount of statistics on $t\bar t$ production already at Run 2 with 139 fb$^{-1}$ of collected data, and there will be much more in the future at its high luminosity upgrade (HL-LHC).
In Refs.~\cite{Afik:2020onf, Fabbrichesi:2021npl, Severi:2021cnj, Afik:2022kwm,Aoude:2022imd}, certain sufficient conditions for entanglement and CHSH violation were obtained, together with a phenomenological exploration of their testability. In Ref.\cite{Severi:2021cnj}, it was concluded that, whereas a verification of entanglement is already feasible from the LHC Run 2 dataset, a significant evidence for the violation of Bell-like inequalities is much harder, even in the high luminosity LHC run.

The purpose of this paper is to re-examine the conditions for entanglement and CHSH-violation in the $t\bar t$ system, 
and devise a refined strategy to experimentally test them. 
To this end, we will make two improvements. First we will consider an upper cut on the velocity of the $t \bar t$ system:
\bea
\beta \equiv \left| \frac{p^z_t + p^z_{\bar t}}{E_t + E_{\bar t}} \right| \leq \beta^{\rm cut} \,,
\label{beta}
\eea
where $p^z$, $E$ are the three-momentum in the direction of the beam axis and the energy, respectively, of the top quark and anti-quark in the laboratory frame. An upper cut on $\beta$ reduces the contribution of $q \bar q$ annihilation processes to $t\bar t$ production with respect to that of gluon fusion: the former are less central due to the different parton distribution functions (PDFs) of valence quarks and sea antiquarks inside the protons.\footnote{A lower cut was proposed in Ref.~\cite{Aguilar-Saavedra:2011dxk} and applied by the ATLAS Collaboration~\cite{ATLAS:2013buu} in the measurement of the $t \bar t$ charge asymmetry, to enhance the $q \bar q$ fraction and measure possible anomalous contributions.}
Therefore, an upper cut on $\beta$ strengthens the entanglement (and thus the CHSH violation) near threshold because the resulting $t\bar t$ system is closer to a spin-singlet (and thus maximally entangled) state due to the Landau-Yang theorem. The second improvement is to design physical observables that allow, by a single direct measurement, to evaluate combinations of spin correlations which enter both the tests of entanglement and the ones of CHSH violation.

The paper is organised as follows. In section \ref{sec:conditions} we re-examine the requirements for entanglement and CHSH violation in the top-antitop system, refining previous sufficient conditions for both, and deriving the sufficient and necessary conditions in physically relevant limits. In section \ref{sec:strategy} we expound the experimental strategy to probe both phenomena at the LHC, in particular we describe the physical observables that give a direct test of both entanglement and CHSH violation. Section \ref{Sec:numerical} is devoted to the numerical simulation of the described strategy, showing the advantage gained thanks to the aforementioned improvements. Finally, in section \ref{sec:discussion} we discuss our results and the future prospects. Additional results are presented in two appendices.

\section{Conditions for entanglement and Bell inequalities }\label{sec:conditions}

An entangled state of two subsystems (Alice and Bob) is by definition a non-separable one, i.e. one that cannot be expressed as $|\psi_1\rangle_A \otimes  |\psi_1\rangle_B$. If the state of the global system is a statistical mixture, described by a density matrix, $\rho$, the separability condition reads
\bea
\rho_{\rm sep} = \sum_n p_n \rho_n^A \otimes \rho_n^B\ ,
\label{rhosep}
\eea
where $p_n$ are classical probabilities and $\sum p_n = 1$. If $\rho$ cannot be expressed as (\ref{rhosep}), then the state is entangled. Mathematically, a necessary and sufficient condition for entanglement in joint systems of two qubits (i.e. each one having a Hilbert space of dimension 2) is provided by the Peres-Horodecki criterion \cite{Peres:1996dw, Horodecki:1997vt}: from the the initial density matrix, $\rho$, a new matrix is constructed by transposing only the indices associated to the Bob (or Alice) Hilbert space. If this partially transposed matrix, say $\rho^{T_2}$, is not a legal density matrix, which in particular means that it has at least one negative eigenvalue, then the $\rho$ matrix corresponds to an entangled state.

Concerning the Bell inequalities, it has been shown that the so-called CHSH inequalities are an optimal version of them for joint systems of two qubits \cite{Fine:1982} when Alice and Bob can measure two different observables each, say $A, A'$ (Alice) and $B, B'$ (Bob), which take (or are assigned to take) two possible values, $\pm 1$. In any local and realistic (`classical') theory, the averages of any such observables must fulfill the CHSH inequality
\bea
\left|\langle AB \rangle  -  \langle AB' \rangle + \langle A'B \rangle + \langle A'B' \rangle \right|\leq 2\ .
\label{CHSH}
\eea
For appropriate choices of the $A, A',B, B'$ observables this inequality can be violated by quantum mechanics in certain entangled $\rho$ states. 

\vspace{0.3cm}
Let us particularise the Peres-Horodecki criterion of entanglement and the CHSH inequalities for a system of two qubits, such as the spin of the $t\bar t$ system. 
An appropriate basis to expand any Hermitian operator acting in the global Hilbert space is provided by the 16 matrices $\left\{\mathbb{1}^A, \sigma_1^A, \sigma_2^A, \sigma_3^A\right\}\otimes\left\{\mathbb{1}^B, \sigma_1^B, \sigma_2^B, \sigma_3^B\right\}$. Hence the density matrix of the joint system, $\rho$, can be written as
\bea
\rho=\frac{1}{4}\left(
\mathbb{1}\otimes \mathbb{1} +\sum_i(B_i^+ \sigma_i\otimes \mathbb{1} + B_i^- \mathbb{1}\otimes \sigma_i) + \sum_{ij}C_{ij} \sigma_i\otimes \sigma_j
\right)\ ,
\label{generalrho}
\eea
where $B_i^\pm, C_{ij}$ are real coefficients and we have dropped the $A, B$ superscripts in the matrices. The $\Tr\ {\rho}=1$ condition is automatically taken into account by the coefficient of the first term. The explicit forms of $\rho, \rho^{T_2}$ read
\footnotesize
\bea
\rho= \frac{1}{4}\left[\begin{array}{cccc} 
1+B_3^++B_3^-+C_{33} & B_1^-+C_{31}-i(B_2^-+C_{32})& B_1^++C_{13}-i(B_2^++C_{23})&
C_{11}-C_{22}-i(C_{12}+C_{21})\cr
B_1^-+C_{31}+i(B_2^-+C_{32})&
1+B_3^+-B_3^--C_{33}&
C_{11}+C_{22}+i(C_{12}-C_{21})&
B_1^+-C_{13}-i(B_2^+-C_{23})\cr
B_1^++C_{13}+i(B_2^++C_{23})&
C_{11}+C_{22}-i(C_{12}-C_{21})&
1-B_3^++B_3^--C_{33}&
B_1^--C_{31}-i(B_2^--C_{32})\cr
C_{11}-C_{22}+i(C_{12}+C_{21})&
B_1^+-C_{13}+i(B_2^+-C_{23})&
B_1^--C_{31}+i(B_2^--C_{32})&
1-B_3^+-B_3^-+C_{33}
\end{array}\right]
\label{rhoexpl}
\eea
\bea
\hspace{-0.25cm}
\rho^{T_2}= \frac{1}{4}\left[\begin{array}{cccc} 
1+B_3^++B_3^-+C_{33} & B_1^-+C_{31}+i(B_2^-+C_{32})& B_1^++C_{13}-i(B_2^++C_{23})&
C_{11}+C_{22}+i(C_{12}-C_{21})\cr
B_1^-+C_{31}-i(B_2^-+C_{32})&
1+B_3^+-B_3^--C_{33}&
C_{11}-C_{22}-i(C_{12}+C_{21})&
B_1^+-C_{13}-i(B_2^+-C_{23})\cr
B_1^++C_{13}+i(B_2^++C_{23})&
C_{11}-C_{22}+i(C_{12}+C_{21})&
1-B_3^++B_3^--C_{33}&
B_1^--C_{31}+i(B_2^--C_{32})\cr
C_{11}+C_{22}-i(C_{12}-C_{21})&
B_1^+-C_{13}+i(B_2^+-C_{23})&
B_1^--C_{31}-i(B_2^--C_{32})&
1-B_3^+-B_3^-+C_{33}
\end{array}\right]
\label{rhoT2expl}
\eea
\normalsize
The eigenvalues of $\rho^{T_2}$ are extremely involved combinations of the various parameters, and so are the necessary and sufficient conditions for entanglement (i.e. the existence of at least one negative eigenvalue). Fortunately, very useful sufficient conditions for entanglement are much easier to obtain, by simply probing the negativity of $v^T \rho^{T_2}v$ for different 4-vectors $v$. In particular, using 
\bea
v= (1,0,0,\pm 1)^T \,,~ (0,1,\pm 1,0)^T
\label{vs}
\eea
we get four completely general {\em sufficient} conditions for entanglement, which can be cast as
\bea
|C_{11}+ C_{22}|&>& 1+ C_{33} \,,
\notag \\
|C_{11}- C_{22}|&>& 1- C_{33} \,,
\label{sigmanegative}
\eea
(it is enough that one of the conditions (\ref{sigmanegative}) is fulfilled to guarantee entanglement).
We remark that the two conditions (\ref{sigmanegative}) are equivalent to the pairs of conditions that can also be obtained for any permutation of the 1,2,3 indices: It can be shown that if neither of (\ref{sigmanegative}) hold, then the analogous conditions with index permutations are not fulfilled either.

One may wonder to which extent
the previous inequalities (\ref{sigmanegative}) are in practice not only sufficient, but also necessary conditions for entanglement. To this end, let us note that one can make some sound approximations in the original density matrix of the $t\bar t$ system (\ref{generalrho}). As shown in Ref.\cite{Bernreuther:2015yna}, $P$ and $CP$ invariance in the $t\bar t$ production leads to $B_i^\pm=0$, $C_{ij}=C_{ji}$, thus reducing the number of parameters from 15 to 6. This is an excellent approximation due to the smallness of the weak corrections to the QCD production mechanism of $t\bar t$ at the LHC.
Besides, using the so-called helicity basis (defined below) as reference system, all the off-diagonal $C_{ij}$ but one (say $C_{12}\simeq C_{21}$), are generated by $P$-odd absorptive parts of the mixed QCD-weak corrections at one-loop, and are very small \cite{Bernreuther:2015yna}. In summary, it is a very good approximation in that basis to keep $C_{11}, C_{22}, C_{33}$ and $C_{12}=C_{21}$ as the only non-vanishing parameters in (\ref{generalrho}). Under this assumption it is possible to extract simple necessary and sufficient conditions for the negativity of $\rho^{T_2}$:
\bea
|C_{11}+ C_{22}|&>& 1+ C_{33} \,, \notag \\
|4C_{12}^2+(C_{11}- C_{22})^2|^{1/2}&>& 1- C_{33} \,,
\label{sigmanegative12}
\eea
(equivalent conditions are obtained by performing  cyclic permutations in the indices). Thus, in the previous approximation, it is sufficient {\em and} necessary that one of the relations (\ref{sigmanegative12}) is fulfilled to guarantee entanglement. It is interesting that the first condition in (\ref{sigmanegative12}) remains unchanged by the inclusion of $C_{12}\neq 0$ and typically represents the most relevant test of entanglement when $C_{11}, C_{22}$ have the same sign and $C_{33}$ is negative, which is precisely the actual situation, as it will be clear in Section~\ref{Sec:numerical} below. Incidentally, in this approximation the conditions for the semipositivity of $\rho$ read
\bea
|C_{11} + C_{22}|&\leq& 1- C_{33} ,
\nonumber\\
|4C_{12}^2 + (C_{11} - C_{22})^2|^{1/2}&\leq& 1+ C_{33} \,
\label{rhopositive12}
\eea
(both conditions required). These relations, together with Eqs.~(\ref{sigmanegative12}), imply that entanglement requires all $C_{ii}\neq 0$.

\vspace{0.3cm}
Let us now turn to the CHSH inequalities Eq.~(\ref{CHSH}). As usual, we consider spin observables for Alice and Bob defined by
\bea
A=a_i \sigma_i^A,\hspace{1cm} A'=a'_i \sigma_i^A, \hspace{1cm} B=b_i \sigma_i^B, \hspace{1cm} B'=b'_i \sigma_i^B ,
\label{observables}
\eea
where, for a moment, we have indicated with a superscript the Alice and Bob operators, and $a_i, a'_i, b_i, b'_i$ are the components of unit-vectors in space in the same basis as the corresponding $\sigma-$matrices. Then the expectation values $\langle AB \rangle, \langle AB' \rangle, \langle A'B \rangle, \langle A'B' \rangle$ for a state described by the density matrix (\ref{generalrho}) read
\bea
\langle AB \rangle={\rm Tr}[\rho (A\otimes B)]=\sum_{ij}C_{ij} a_ib_j \ ,
\eea
with analogous expressions obtained by replacing $A\rightarrow A'$, $a_i\rightarrow a'_i$; $B\rightarrow B'$, $b_j\rightarrow b'_j$.
Hence the CHSH inequality (\ref{CHSH}) can be expressed as
\bea
\left|\vec{a}\cdot C (\vec{b}-\vec{b'})
\ +\ 
\vec{a'}\cdot C (\vec{b}+\vec{b'}) \right|\leq 2\ .
\label{CHSH-abC}
\eea
We are interested in choices of $\vec{a},\vec{a'},\vec{b},\vec{b'}$ that violate (\ref{CHSH-abC}) in the strongest possible way.
In this sense, it was shown in Ref.\cite{Horodecki:1995} that the maximum
of the l.h.s. of (\ref{CHSH-abC}) is given by
\bea
\max_{{a},{a'},{b},{b'}}\left|\vec{a}\cdot C (\vec{b}-\vec{b'})
\ +\ 
\vec{a'}\cdot C (\vec{b}+\vec{b'}) \right|=2\sqrt{\lambda+\lambda'}\ ,
\label{CHSH-abC_max}
\eea
where $\lambda,\lambda'$ are the largest eigenvalues of $C^T C$. This corresponds to choosing $\vec{b}\pm\vec{b'}$ proportional to the associated eigenvectors with appropriate coefficients, and $\vec{a},\vec{a'}$ as the unit-vectors proportional to 
$C(\vec{b}-\vec{b'}), C(\vec{b}+\vec{b'})$, respectively. Then the {\em violation} of any CHSH inequality requires
\bea
\lambda+\lambda'>1\ ,
\label{optimalCHSH}
\eea
which thus represents, in principle, the optimal test for CHSH violation.
However, as discussed in Ref.\cite{Severi:2021cnj}, the strategy of relying on the largest eigenvalues of $C^T C$, Eq.~(\ref{optimalCHSH}), to test CHSH violation \cite{Fabbrichesi:2021npl}
is in practice problematic and likely biased, due to the uncertainties involved in the measurement.
A more conservative approach consists in choosing $\vec{a},\vec{a'}, \vec{b},\vec{b'}$ in such a way that the l.h.s. of (\ref{CHSH-abC}) is still potentially larger than 2, but far simpler than (\ref{CHSH-abC_max}) and with much more transparent physical meaning. In particular, for any pair of indices $i\neq j$ we can choose
\bea
a_k=\delta_{ki},&\hspace{1cm}& a'_k=\delta_{kj},
\nonumber\\
b_i=-b'_i=\pm \frac{1}{\sqrt{2}},
&\hspace{1cm}&
b_j=b'_j=\pm \frac{1}{\sqrt{2}},
\hspace{1cm}
b_{k\neq i,j}=0
\eea
(uncorrelated signs). Using these two sets of vectors the violation of CHSH inequalities reads
\bea
|C_{ii}\pm C_{jj}| > \sqrt{2}\ .
\label{CHSHij}
\eea
Then, of course, the strategy is to choose $i,j$ and the $\pm$ sign to maximise $|C_{ii}\pm C_{jj}|$ (the optimal choice depending on their moduli and relative sign).
It is interesting to note that, although Eq.~(\ref{CHSHij}) involves just two diagonal entries of $C$, the third one must also be non-vanishing. Actually, it occurs in such a way that 
at least one of the sufficient conditions for entanglement (\ref{sigmanegative}) is fulfilled.\footnote{This can be easily shown, by taking into account that $\rho$ must be a positive-semidefinite matrix. In particular, the positivity of $v^T \rho v$ for the vectors (\ref{vs}) implies the necessary conditions
$|C_{11} \pm C_{22}|\leq 1\mp C_{33}$, valid for any permutation of the $1,2,3$ indices.}

Similarly to our previous discussion on the entanglement test, one may wonder to which extent expression (\ref{CHSHij}) is close in practice to the optimal value (\ref{CHSH-abC_max}). As discussed above, an excellent approximation for $C$ is to assume $C_{ij}=C_{ji}$, so $\lambda, \lambda'$ are the largest eigenvalues of $C^2$. Then, from the eigenvalue interlacing theorem, for any $i\neq j$
\bea
2\sqrt{\lambda+\lambda'}\ \geq\  2\sqrt{(C^2)_{ii} + (C^2)_{jj}}\ \geq\ 
2\sqrt{(C_{ii})^2 + (C_{jj})^2}\ \geq\ 
\sqrt{2}|C_{ii}\pm C_{jj}|\ .
\label{hierarchy}
\eea
Note that the violation of the CHSH inequality (\ref{CHSHij}) implies condition (\ref{optimalCHSH}), as expected, but not the other way around. Now, the first two inequalities in (\ref{hierarchy}) are usually close to the equality due to the smallness of the off-diagonal $C-$entries in the bases we use, as it will be clear below in Section~\ref{sec:4}. The third inequality in (\ref{hierarchy}) is close to the identity, as long as $(|C_{ii}|-|C_{jj}|)^2\ll (|C_{ii}|+|C_{jj}|)^2$, which is also the typical case 
once we select the two largest $|C_{ii}|,\ |C_{jj}|$. Anyway, if desired, one  can refine the CHSH-violation condition (\ref{CHSHij}) with this correction:
\bea
|C_{ii}|+|C_{jj}|+
\frac{1}{2}\frac{(|C_{ii}|-|C_{jj}|)^2}
{|C_{ii}|+ |C_{jj}|}
> \sqrt{2}\ ,
\label{CHSHijref}
\eea
which, of course, is easier to satisfy than (\ref{CHSHij}).

\section{Experimental strategy}
\label{sec:strategy}

In this section we use
the charged leptons from the top and antitop semileptonic decays $t \bar t \to \ell^+ \nu b \, \ell^- \bar \nu \bar b$ to analyse the spin properties of the mother particles; but, in order to keep the formalism general, we first consider arbitrary decay products `$a$' from the top quark and `$b$' from the anti-quark. Their normalised three-momenta, expressed in polar coordinates, in the respective top quark and antiquark rest frames, read
\begin{align}
& \hat{p}_{a} = (\sin \theta_a \cos \varphi_a,\sin \theta_a \sin \varphi_a,\cos \theta_a )^T \,, \notag \\
& \hat{p}_{b} = (\sin \theta_b \cos \varphi_b,\sin \theta_b \sin \varphi_b,\cos \theta_b )^T \,.
\label{pipj}
\end{align}
where we use he same $R^3$ basis $(\hat x,\hat y, \hat z)$ in the top quark and antiquark rest frames. The decay density matrices for the quarks read~\cite{Boudjema:2009fz}
\begin{equation}
\Gamma_a = \frac{1}{2} \left( \begin{array}{cc} 
1 + \alpha_a \cos \theta_a & \alpha_a \sin \theta_a e^{i \varphi_a} \\
\alpha_a \sin \theta_a e^{-i \varphi_a} & 1 - \alpha_a \cos \theta_a
\end{array} \right) \,,
\end{equation}
with an analogous expression for $\Gamma_b$ with the replacement $a \to b$.
The constants $\alpha$ satisfy $|\alpha| \leq 1$ and are called `spin analysing power' of the corresponding top (anti-)quark decay product. For the charged leptons $\alpha_{\ell^+} = - \alpha_{\ell^-} = 1$ at the tree level, with  next-to-leading order corrections at the permille level~\cite{Brandenburg:2002xr}.
The joint decay distribution is proportional to the sum~\cite{Rahaman:2021fcz}
\begin{equation}
\sum_{\lambda_t \lambda'_t \lambda_{\bar t} \lambda'_{\bar t} = \pm 1/2} \rho(\lambda_t, \lambda'_t, \lambda_{\bar t}, \lambda'_{\bar t}) \Gamma_a(\lambda_t, \lambda'_t) \Gamma_b(\lambda_{\bar t}, \lambda'_{\bar t}) \,.
\end{equation}
Omitting for brevity the terms proportional to $B_i^\pm$, which do not contribute to the observables considered,
the quadruple differential distribution reads 
\begin{eqnarray}
\frac{1}{\sigma} \frac{d\sigma}{d\Omega_a d\Omega_b} & = &
\frac{1}{(4\pi)^2} \left[
1 + \alpha_a \alpha_b \sin \theta_a \sin \theta_b 
\left( C_{11} \cos \varphi_a \cos \varphi_b + C_{22} \sin \varphi_a \sin \varphi_b \right)
\right. \notag \\
& & + \alpha_a \alpha_b \sin \theta_a \sin \theta_b
\left( C_{12} \cos \varphi_a  \sin \varphi_b + C_{21} \sin \varphi_a \cos \varphi_b \right)
\notag \\
& & + \alpha_a \alpha_b \left( C_{13} \sin \theta_a \cos \varphi_a \cos \theta_b
+ C_{31} \cos \theta_a \sin \theta_b \cos \varphi_b \right )
\notag \\
& & + \alpha_a \alpha_b \left( C_{23} \sin \theta_a \sin \varphi_a \cos \theta_b + C_{32} \cos \theta_a \sin \theta_b \sin \varphi_b \right)
\notag \\
& & \left. + \alpha_a \alpha_b  C_{33} \cos \theta_a \cos \theta_b \right] \,,
\label{ec:dist4D-1}
\end{eqnarray}
with $d\Omega_a = d\cos \theta_a d\varphi_a$ and likewise for $d\Omega_b$. 
Using this distribution, we want to design observables that directly test entanglement and violation of the CHSH inequalities.

For completeness, we point out that the coefficients can be individually measured from two-dimensional distributions or from forward-backward asymmetries, without the need of a fit to the four-dimensional distribution (\ref{ec:dist4D-1}). Let us relabel for convenience the polar angles $\theta_{a,b}$ in (\ref{pipj}) as $\theta_{a,b}^3$, and introduce the polar angles $\theta_{a,b}^{1}$, $\theta_{a,b}^{2}$ between $\hat p_{a,b}$ and the $\hat x$, $\hat y$ axes, respectively. Then, for $i,j=1,2,3$ we have the distributions~\cite{Bernreuther:2015yna}
\begin{equation}
\frac{1}{\sigma}\frac{d\sigma}{d\cos \theta_a^i d\cos \theta_b^j} = 
\frac{1}{4} \left(1 + \alpha_a \alpha_b C_{ij} \cos \theta_a^i \cos \theta_b^j \right) \,.
\end{equation}
Hence, $C_{ij}$ can be extracted by a fit to the above distributions, properly corrected for detector effects. Alternatively, one can write forward-backward asymmetries
\begin{equation}
A_{ij} = \frac{N(\cos \theta_a^i \cos \theta_b^j > 0) - N(\cos \theta_a^i \cos \theta_b^j < 0)}
{N(\cos \theta_a^i \cos \theta_b^js > 0) + N(\cos \theta_a^i \cos \theta_b^j < 0)} = \frac{1}{4} \alpha_a \alpha_b C_{ij} \,,
\label{ec:Akl}
\end{equation}
with $N(\cdot)$ standing for the number of events. These asymmetries allow to extract the coefficients, assuming the Standard Model value for $\alpha_a \alpha_b$.

\subsection{Observables for entanglement}

A sufficient set of conditions for entanglement is given in (\ref{sigmanegative}). Depending on the sign of the sums in the absolute value, the quantities to test are
\begin{align}
& C_{11} + C_{22} - C_{33} > 1 \,, \notag \\
& -C_{11} - C_{22} - C_{33} > 1 \,, \notag \\
& C_{11} - C_{22} + C_{33} > 1 \,, \notag \\
& -C_{11} + C_{22} + C_{33} > 1 \,.
\label{ec:3Csumsexp}
\end{align}
In the second line we can identify the sum $C_{11} + C_{22} + C_{33}$. There is a well-known observable that can measure this sum~\cite{Bernreuther:2010ny,Bernreuther:2004jv,Bernreuther:2013aga,Bernreuther:2015yna}; 
namely, the one-dimensional kinematical distribution in terms of the cosine of the `opening angle' $\cos \theta_{ab} \equiv \hat p_a \cdot \hat p_b$, with  $\hat p_a, \hat p_b$ defined in (\ref{pipj}),
\begin{equation}
\frac{1}{\sigma} \frac{d\sigma}{d\cos \theta_{ab}} = \frac{1}{2} \left( 1 + \alpha_a \alpha_b D \cos \theta_{ab} \right) \,,
\end{equation}
where the $D$ coefficient is related to the trace of the $C$ matrix by\cite{Bernreuther:2004jv}\footnote{Our sign convention is different from that in Refs.~\cite{Bernreuther:2010ny,Bernreuther:2004jv,Bernreuther:2013aga,Bernreuther:2015yna} because in the definition of the $\rho$ matrix (\ref{generalrho}), as well as for the coordinate systems in the $t,\bar t$ rest frames, we use the same reference system, therefore introducing a sign difference in the $C$ coefficients.}
\begin{equation}
D = \frac{1}{3} (C_{11} + C_{22} + C_{33}) \,.
\label{ec:Drel}
\end{equation}
This coefficient can be experimentally measured either by a fit to the $\cos \theta_{ab}$ distribution, or by a forward-backward asymmetry in $\cos \theta_{ab}$.
The rest of equations in (\ref{ec:3Csumsexp}) involve two coefficients with the same sign and the third with opposite sign. It is easy to design an observable for these sums. The differential distribution (\ref{ec:dist4D-1}) can be written in compact form as
\begin{equation}
\frac{1}{\sigma} \frac{d\sigma}{d\Omega_a d\Omega_b} = \frac{1}{(4\pi)^2} \left[ 1 + \alpha_a \alpha_b \hat p_a^T C \hat p_b \right] \,.
\label{ec:dist4D-C}
\end{equation}
Let us focus for definiteness on the sum in the first line of (\ref{ec:3Csumsexp}), $C_{11} + C_{22} - C_{33}$. By introducing a reflection in the $(x_1,x_2)$ plane of the top anti-quark rest frame, $P_3 = \text{diag}(1,1,-1)$,  the angular-dependent term in (\ref{ec:dist4D-C}) can be rewritten as
\begin{equation}
\hat p_a^T C \hat p_b = \hat p_a^T C P_3^2 \hat p_b = \hat p_a^T (C P_3) (P_3 \hat p_b) \,, 
\end{equation}
with
\begin{align}
& (P_3 \hat p_b) = (\sin \theta_b \cos \varphi_b,\sin \theta_b \sin \varphi_b,-\cos \theta_b )^T \,.
\label{ec:Pjmirror}
\end{align}
Therefore, the kinematical distribution of the angle $\theta_{ab}'$, defined by $\cos \theta_{ij}' = \hat p_a \cdot P_3\hat p_b$,
\begin{equation}
\frac{1}{\sigma} \frac{d\sigma}{d\cos \theta'_{ab}} = \frac{1}{2} \left( 1 + \alpha_a \alpha_b D_3 \cos \theta'_{ab} \right) \,,
\end{equation}
measures the trace of the matrix $(C P_3)$, that is,
\begin{equation}
D_3 = \frac{1}{3} (C_{11} + C_{22} - C_{33}) \,.
\label{ec:D3rel}
\end{equation}
This dependence is confirmed by directly calculating $\langle \cos \theta'_{ab} \rangle$ using (\ref{ec:dist4D-1}). Likewise as for $D$, the $D_3$ coefficient can be measured either with a fit to the $\cos \theta'_{ab}$ distribution
or by a forward-backward asymmetry in $\cos \theta'_{ab}$.

\subsection{Observables for CHSH inequalities}

We now want to design observables that can directly measure sums and differences $C_{ii} \pm C_{jj}$ involved in the relations for CHSH violation (\ref{CHSHij}) or (\ref{CHSHijref}).  Let us define the azimuthal angle differences
\begin{align}
& \varphi_+ = \frac{1}{2}(\varphi_a + \varphi_b) \,, \quad \varphi_- = \frac{1}{2} (\varphi_a - \varphi_b ) \,.
\end{align}
Changing variables in (\ref{ec:dist4D-1}) to these angles, we arrive at
\begin{align}
 \frac{1}{\sigma} \frac{d\sigma}{d\cos \theta_a d\cos \theta_b d\varphi_+ d\varphi_-} = \notag \\
& \frac{2}{(4\pi)^2} \left\{
1 + \alpha_a \alpha_b \sin \theta_a \sin \theta_b \left[ 
\frac{C_{11} + C_{22}}{2} \cos 2 \varphi_-
+ \frac{C_{11} - C_{22}}{2} \cos 2 \varphi_+ \right. \right. \notag \\
& \left. + \frac{C_{12} + C_{21}}{2} \sin 2\varphi_+ + \frac{C_{21}-C_{12}}{2} \sin 2\varphi_- \right]
+ \alpha_a \alpha_b C_{33} \cos \theta_a \cos \theta_b \notag \\
& + \alpha_a \alpha_b \sin \theta_a \cos \theta_b \left[
C_{13} \cos(\varphi_+ + \varphi_-) + C_{23} \sin(\varphi_+ + \varphi_-) \right] \notag \\
& + \alpha_a \alpha_b \cos \theta_a \sin \theta_b \left[
C_{31} \cos(\varphi_+ - \varphi_-) + C_{32} \sin(\varphi_+ - \varphi_-) \right] \,.
\label{ec:dist4D-2}
\end{align}
Since we are only interested in the sums in the first line of (\ref{ec:dist4D-2}), we can integrate over $\cos \theta_{a,b}$ to obtain
\begin{eqnarray}
\frac{1}{\sigma} \frac{d\sigma}{d\varphi_+ d\varphi_-} & = & \frac{1}{2\pi^2} + \frac{\alpha_a \alpha_b}{32} \left[
\frac{C_{11} + C_{22}}{2} \cos 2 \varphi_- + \frac{C_{11} - C_{22}}{2} \cos 2 \varphi_+ \right. \notag \\
& & \left. + \frac{C_{12} + C_{21}}{2} \sin 2 \varphi_+ + \frac{C_{21}-C_{12}}{2} \sin 2 \varphi_- \right] \,.
\end{eqnarray}
Integration over $\varphi_+$ and $\varphi_-$ using the integration measure $g_+ = \text{sign} \cos 2 \varphi_-$, yields the asymmetry
\begin{equation}
A_+ = \frac{1}{\sigma} \int \frac{d\sigma}{d\varphi_+ d\varphi_-} g_+ \, d\varphi_+ d\varphi_- = \frac{\pi}{16} \alpha_a \alpha_b (C_{11} + C_{22}) \,,
\label{ec:Aplus}
\end{equation}
whereas if we use the integration measure $g_- = \text{sign} \cos 2 \varphi_+$
we obtain the asymmetry
\begin{equation}
A_- = \frac{1}{\sigma} \int \frac{d\sigma}{d\varphi_+ d\varphi_-} g_- \, d\varphi_+ d\varphi_- = \frac{\pi}{16} \alpha_a \alpha_b (C_{11} - C_{22}) \,.
\label{ec:Aminus}
\end{equation}
Experimentally, these asymmetries are measured by simply counting events with $\cos (\varphi_a \mp \varphi_b) > 0$ and $\cos (\varphi_a \mp \varphi_b) < 0$, that is,
\begin{equation}
A_\pm = \frac{N(\cos (\varphi_a \mp \varphi_b) > 0) - N(\cos (\varphi_a \mp \varphi_b) < 0)}{N(\cos (\varphi_a \mp \varphi_b) > 0) + N(\cos (\varphi_a \mp \varphi_b) < 0)} \,.
\end{equation}

\section{Numerical results}\label{Sec:numerical}
\label{sec:4}

We use the Monte Carlo generator {\scshape MadGraph\_aMC@NLO}~\cite{Alwall:2014hca} to obtain our predictions for $t \bar t$ production at the leading order in QCD, using NNPDF 3.0~\cite{NNPDF:2014otw} PDFs with factorisation and renormalisation scales equal to the average transverse mass, $Q = 1/2 [(m_t^2 + p_{T t}^2)^{1/2} + (m_t^2 + p_{T \bar t}^2)^{1/2} ]$, with $p_T$ the transverse momentum in the usual notation. The dilepton decay channel $t \bar t \to \ell^+ \nu b \ell^- \nu b$ is selected, with $\ell = e,\mu$. In order to study the dependence on the $t\bar t$ invariant mass $\mtt$ and the top scattering angle in the center of mass frame $\thCM$, we generate samples of $t \bar t$ events in 50 GeV slices of $\mtt$ between $[300,350]$ and $[750,800]$ GeV, and 100 GeV slices between $[0.8,0.9]$ and $[1.2,1.3]$ TeV, totaling $3.4 \times 10^7$ events. Three additional samples of $5 \times 10^6$ events each are generated, with an upper cut $\mtt \leq 360$ GeV, with a lower cut $\mtt \geq 1$ TeV and without cuts on $\mtt$. 

In our Monte Carlo calculations we work at the parton level and do not set any cut on the transverse momenta of the visible particles (charged leptons and $b$-quark jets) nor missing energy, in order to obtain predictions for the full phase space. An experimental analysis involves some loose lower cuts on the transverse momenta of the visible particles, and some upper cuts on their rapidities, typically $|\eta| \leq 2.5$. A veto is also placed on events where the two leptons have the same flavour and invariant mass $m_{\ell \ell} \sim M_Z$, in order to suppress the Drell-Yan background. In addition, the momenta of the top quark and anti-quark have to be reconstructed from those of the observed particles and the missing energy. Therefore, the detector-level observables and/or distributions measured have to be unfolded to take into account acceptance, detector and reconstruction effects, and recover their parton-level values. This is a well-established procedure (see for example Refs.~\cite{ATLAS:2016bac,CMS:2019nrx}) and, because we are only interested on the improvement of the statistical sensitivity, we do not address these issues. Instead, we use the parton-level information (momenta of all the particles, including top quarks) to determine the theoretical predictions of the observables considered.

In this work we mainly use as reference system to express the $t\bar t$ density matrix (\ref{generalrho}) the helicity basis, with vectors $(\hat r,\hat n,\hat k)$ defined as
\begin{itemize}
\item K-axis (helicity): $\hat k$ is a normalised vector in the direction of the top quark three-momentum in the $t \bar t$ rest frame.
\item R-axis: $\hat r$ is in the production plane and defined as $\hat r = \mathrm{sign}(y_p) (\hat p_p - y_p \hat k)/r_p$, with $\hat p_p = (0,0,1)$ the direction of one proton in the laboratory frame, $y_p = \hat k \cdot \hat p_p$, $r_p = (1-y_p^2)^{1/2}$. The definition for $\hat r$ is the same if we use the direction of the other proton $- \hat p_p$.
\item N-axis: $\hat n = \hat k \times \hat r$ is orthogonal to the production plane and can also be written as $\hat n = \mathrm{sign}(y_p) (\hat p_p \times \hat k)/r_p$, which again is independent of the proton choice.
\end{itemize}
We point out that, unlike other works~\cite{Bernreuther:2015yna,Fabbrichesi:2021npl} we use the same basis for the top quark and anti-quark. For the test of the CHSH inequalities at threshold we use a fixed beamline basis $(\hat x,\hat y,\hat z)$ with $\hat x = (1,0,0)$, $\hat y = (0,1,0)$, $\hat z = (0,0,1)$.

\begin{figure*}[htb]
\begin{center}
\begin{tabular}{cc}
\includegraphics[width=5.5cm,clip=]{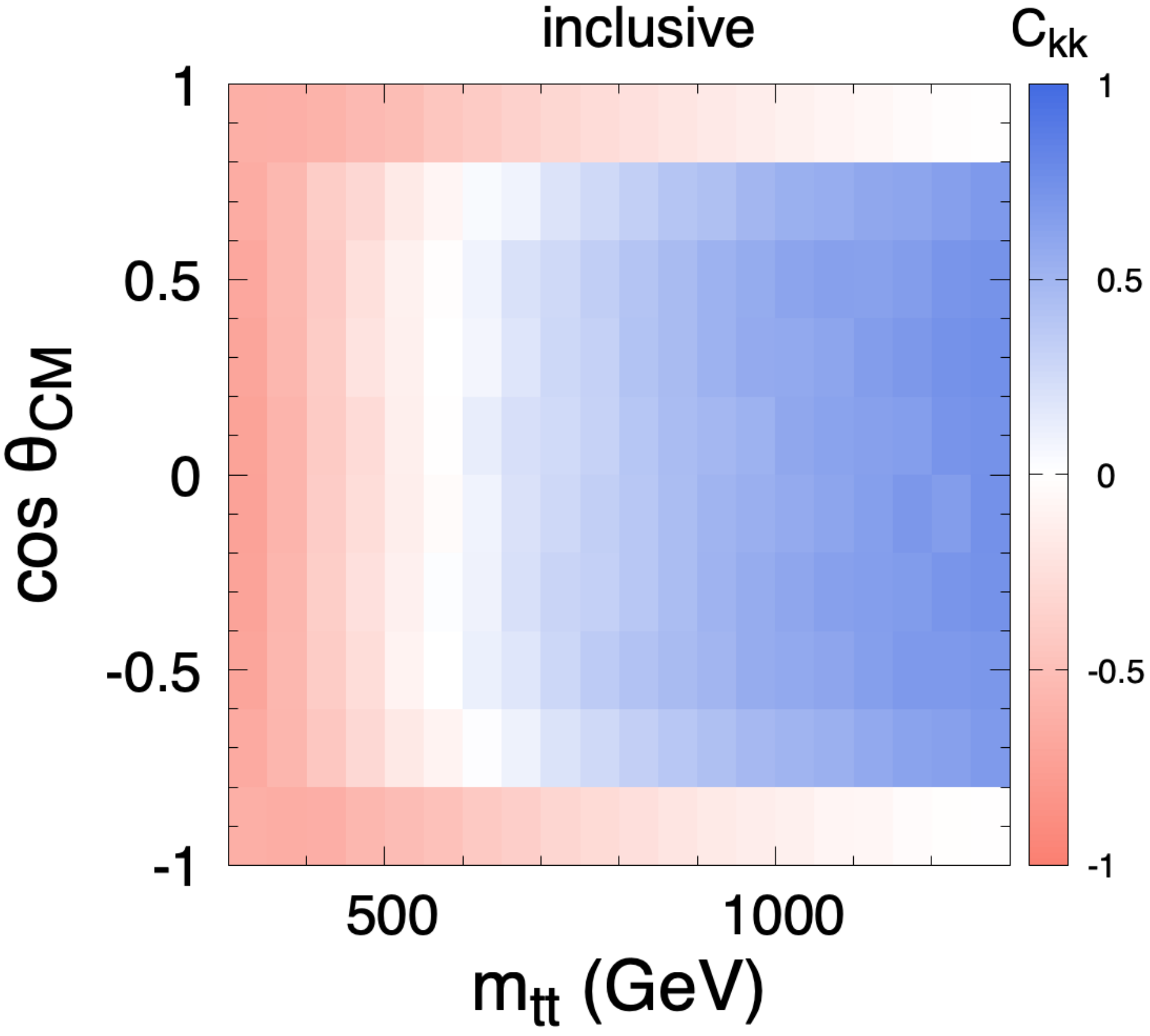}
& \includegraphics[width=5.5cm,clip=]{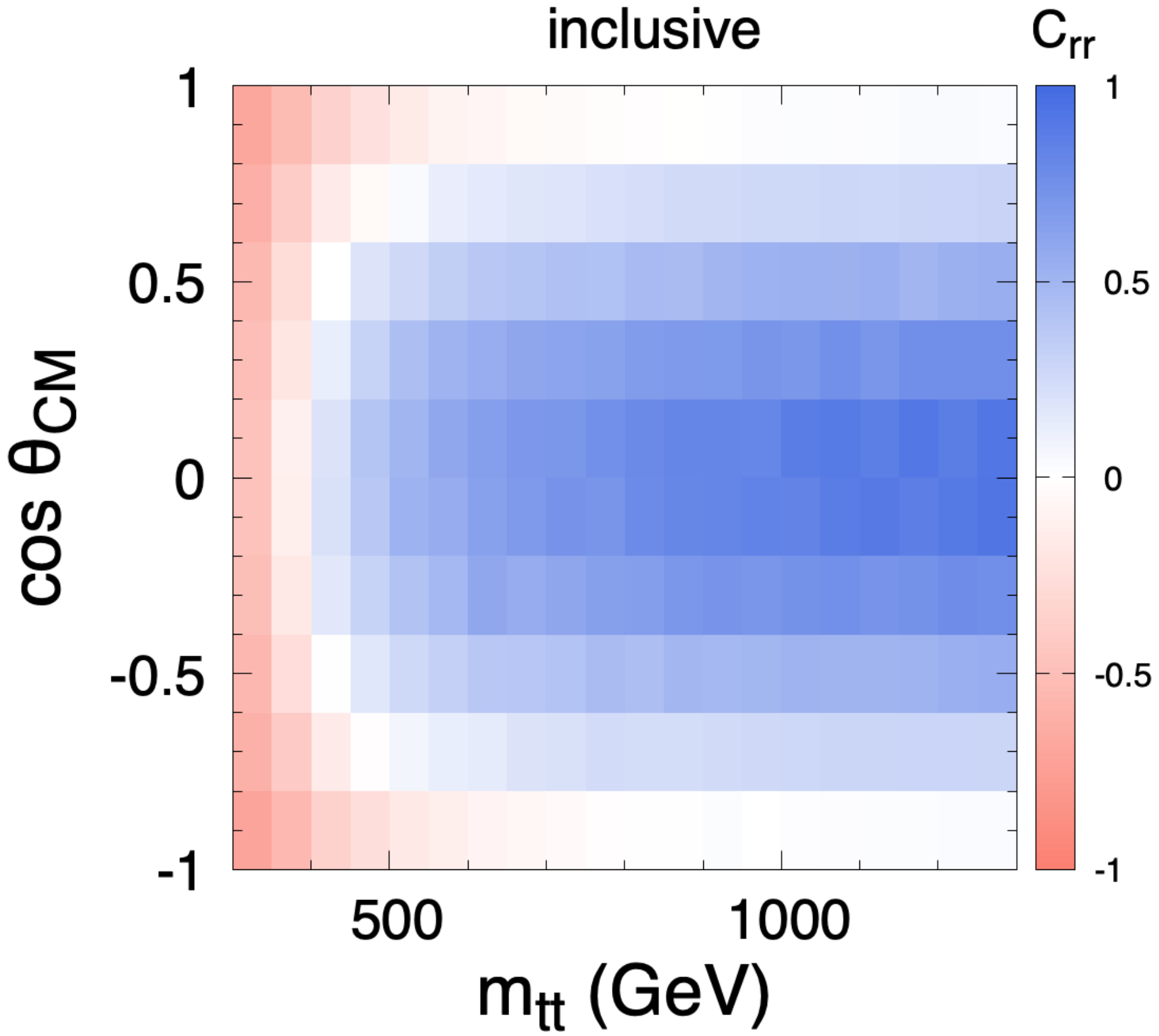} \\
\includegraphics[width=5.5cm,clip=]{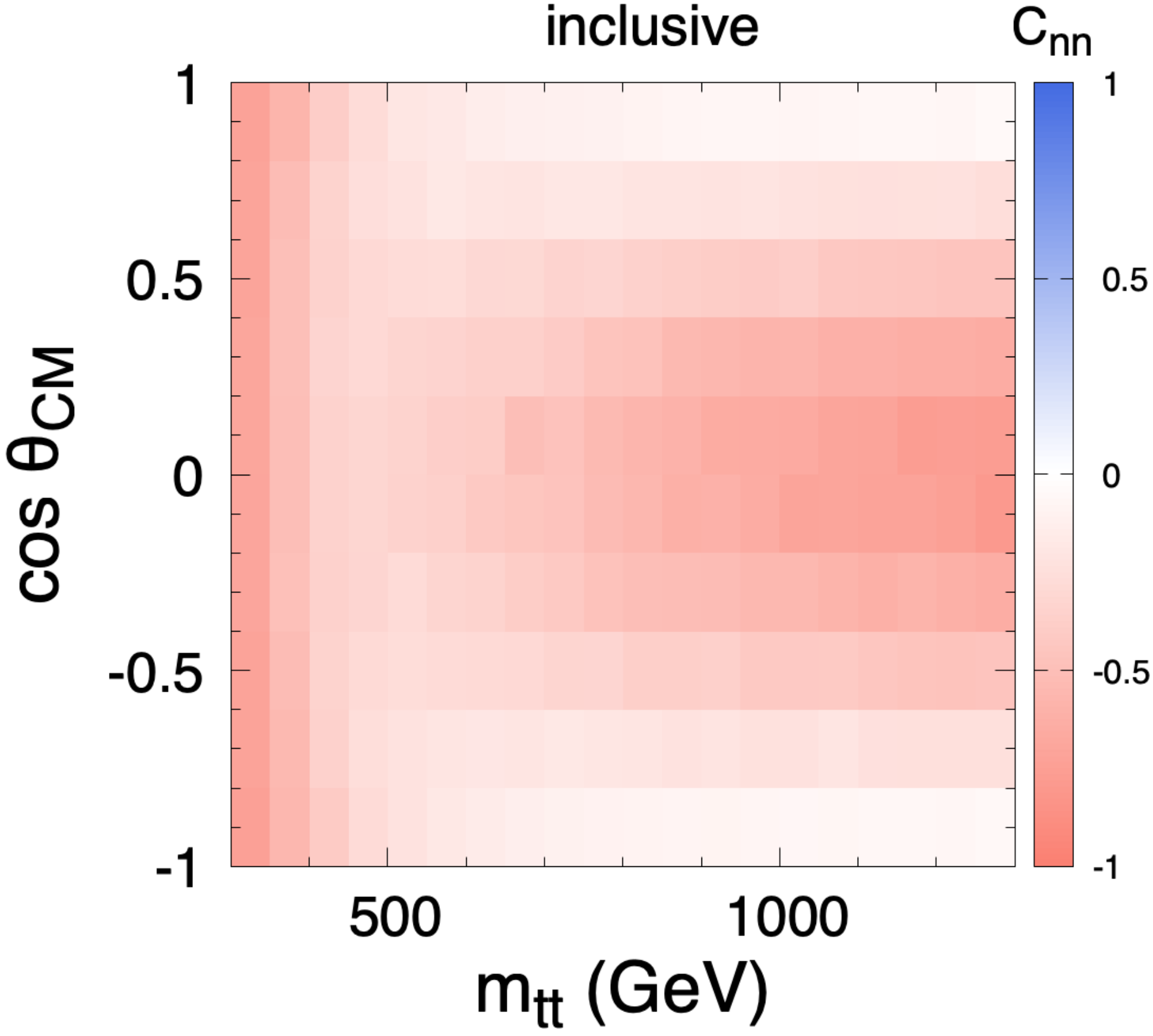} &
\includegraphics[width=5.5cm,clip=]{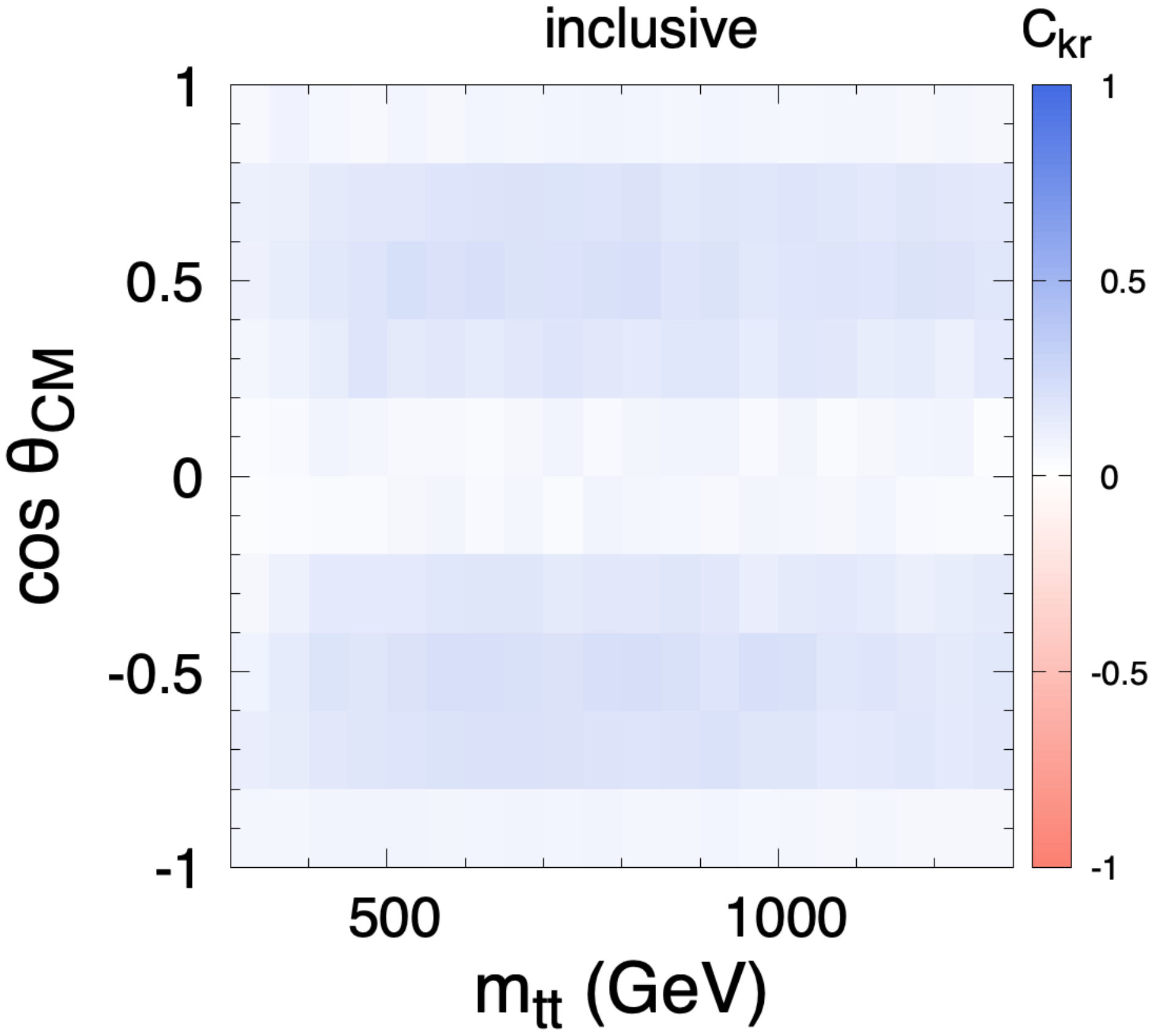} 
\end{tabular}
\caption{Dependence on $\mtt$ and $\thCM$ of the spin correlation coefficients $C_{kk}$, $C_{rr}$, $C_{nn}$ and $C_{kr}$, at the parton level without kinematical cuts.}
\label{fig:C}
\end{center}
\end{figure*}

In order to motivate the choice of observables in the following subsections,
we present in Fig.~\ref{fig:C} the dependence on $\mtt$ and $\thCM$ of the diagonal spin correlation coefficients $C_{kk}$, $C_{rr}$, $C_{nn}$ in the helicity basis, as well as the largest off-diagonal coefficient $C_{kr} = C_{rk}$. The integrated values of the spin correlation coefficients are $C_{kk} = -0.346$, $C_{rr} = -0.021$, $C_{nn} = -0.334$, $C_{kr} = 0.109$, close to the NLO values~\cite{Bernreuther:2015yna}. Note that $C_{nn} \leq 0$ across all the range studied, while for $C_{kk}$ and $C_{rr}$ there can be large correlations of either sign. In particular, although in average $C_{rr}$ is the smallest correlation, when the values in specific regions are considered, $C_{kr}$ is by far the smallest one.
All the figures are approximately symmetric in $\cosCM$ but we prefer to keep the $[-1,1]$ range of variation of this variable in order to have a visual estimation of the statistical uncertainty of our Monte Carlo results. We note that for simplicity, throughout this work we calculate the individual spin correlation coefficients from forward-backward asymmetries of the type (\ref{ec:Akl}).
In the next subsections we study in more detail how the use of an upper cut on $\mybeta$ improves the observability of quantum correlations and the violation of the CHSH inequalities.

\subsection{Observation of entanglement}
\label{sec:4.1}

In view that in the helicity basis $C_{nn} \leq 0$ and $C_{kk},\ C_{rr}$ have typically the same sign, the most useful entanglement condition among Eqs.~(\ref{ec:3Csumsexp}) is
\begin{equation}
E \equiv |C_{kk} + C_{rr}| - C_{nn} -1 > 0 \,.
\label{ec:E1}
\end{equation}
For brevity, we label the entanglement indicator in the l.h.s. of the inequality (\ref{ec:E1}) as $E$. Near threshold both $C_{kk}$ and $C_{rr}$ are negative (see Fig.~\ref{fig:C}), therefore $E = -C_{kk} - C_{rr} - C_{nn} - 1 = - 3D -1$, c.f. (\ref{ec:Drel}). For boosted central tops $C_{kk},C_{rr} > 0$, and $E = C_{kk} + C_{rr} - C_{nn} - 1 = 3 D_3 -1$, as follows from (\ref{ec:D3rel}) by choosing the third axis in the $\hat n$ direction.

\begin{figure}[htb]
\begin{center}
\begin{tabular}{ccc}
\includegraphics[width=5.5cm,clip=]{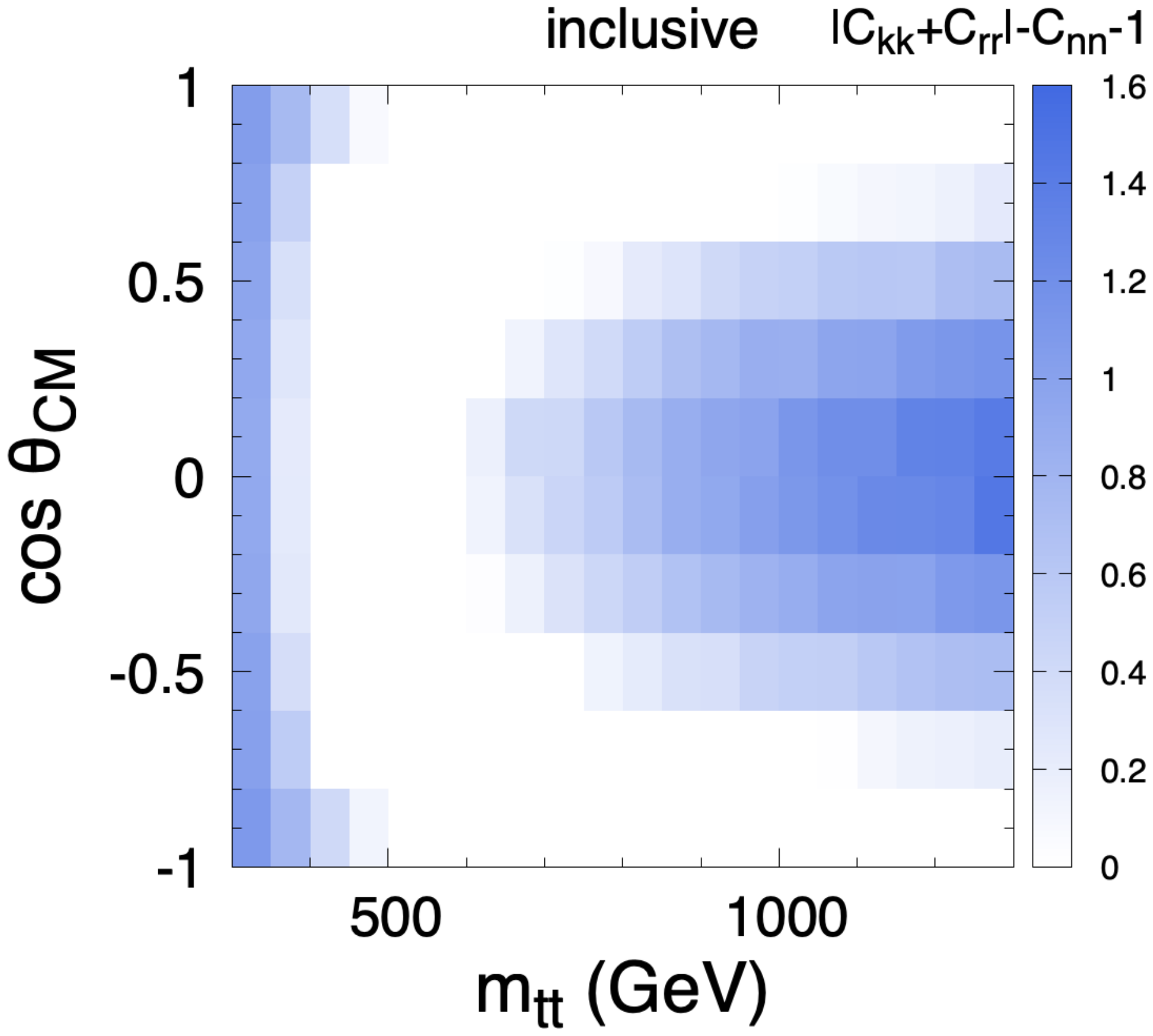} & \quad &
\includegraphics[width=5.5cm,clip=]{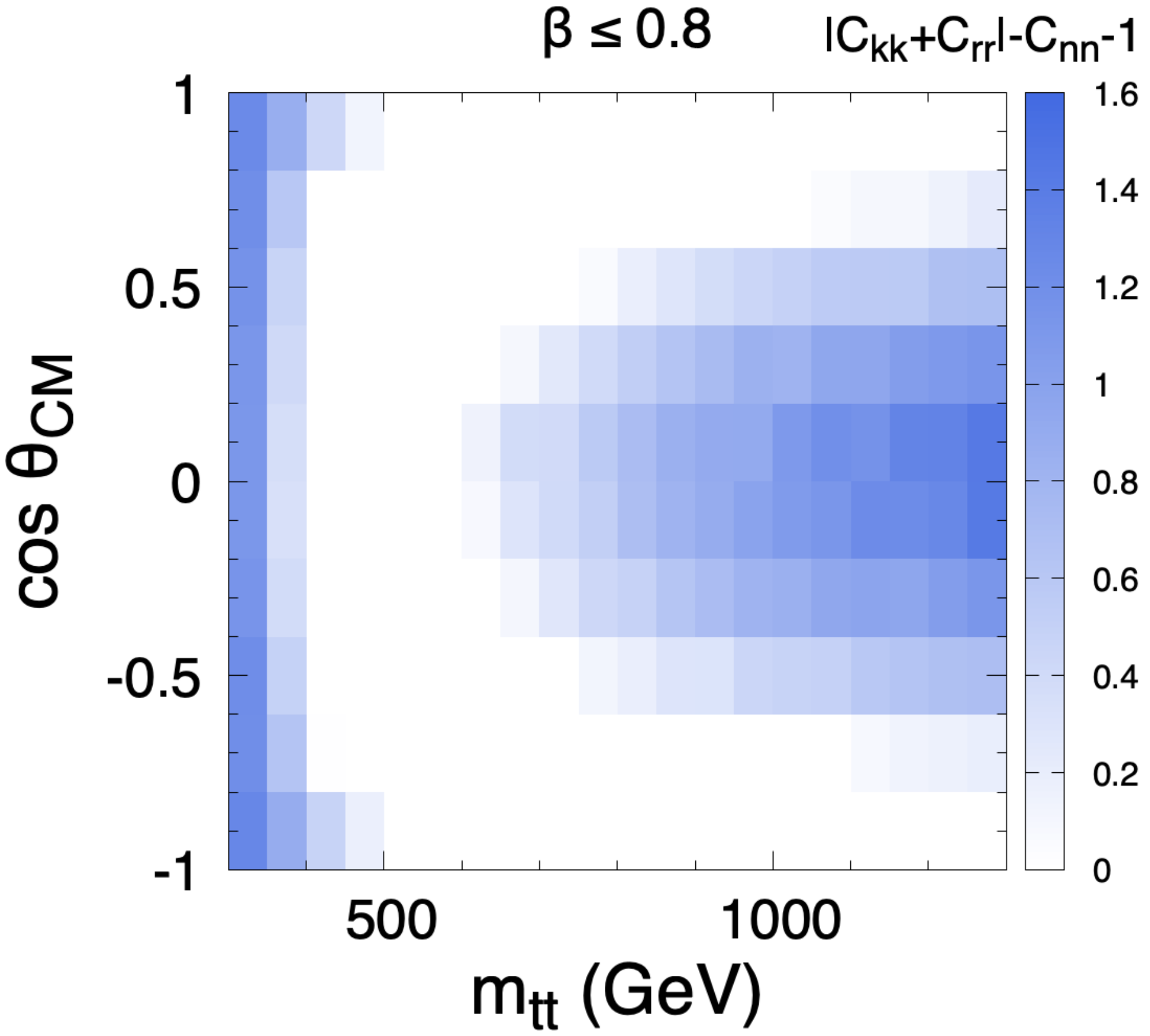}
\end{tabular}
\caption{Dependence of the entanglement indicator $E$ in (\ref{ec:E1}) on $\mtt$ and $\thCM$.}
\label{fig:E1}
\end{center}
\end{figure}

As discussed below Eq.~(\ref{beta}), an upper cut on $\mybeta$ potentially enhances the entanglement near threshold. In Fig.~\ref{fig:E1}, left panel, we plot $E$ (evaluated from the individual values of $C_{ii}$) as a function of $\mtt$ and $\thCM$ without applying any kinematical cut, whereas in the right panel we require $\mybeta \leq 0.8$. The enhancement is notable near threshold, while the requirement on $\beta$ has little effect at the boosted, central region.
Of course, in order to optimise the sensitivity to entanglement one has to consider the uncertainty in the measurement in each case. In this work we do not consider systematic uncertainties, which cannot be addressed without a full detector simulation, a complete reconstruction of the $t \bar t$ kinematics, and unfolding to parton-level. Instead, we just focus on statistical uncertainties. These basically depend on the size of the signal sample after event selection. In this work we assume an event selection and reconstruction efficiency of $0.12$, the average value found in Ref.~\cite{Severi:2021cnj}. For comparison, using a typical event selection and a $t \bar t$ kinematical reconstruction, an efficiency of 0.17 was found in Ref.~\cite{Aguilar-Saavedra:2021ngj} using {\scshape Pythia}~\cite{Sjostrand:2007gs} for hadronisation and showering and {\scshape Delphes}~\cite{deFavereau:2013fsa} for a fast simulation of the detector. This efficiency is used to obtain the number of events, given the $t \bar t \to \ell^+ \nu b \ell^- \bar \nu \bar b$ cross section in the phase space region considered and the luminosity. 

We devise three sets of simple kinematical cuts to enhance the observability of $E$ near threshold and at the boosted regime; in the former case, with and without the use of $\beta$, to illustrate the improvement.
Since statistical uncertainties are proportional to $1/\sqrt \sigma$, we use the figure of merit
\begin{equation}
S_E = E \times \sqrt \sigma
\end{equation}
to guide the selection of the kinematical cuts, collected in Table~\ref{tab:cutE}, together with the tree-level cross section after these cuts.
The last column corresponds to the number of events with a luminosity $L = 139~\text{fb}^{-1}$ collected at the LHC Run 2, assuming a $K$ factor of 1.8 to normalise the total cross section to next-to-next-to-leading order~\cite{Czakon:2011xx} and using a selection and reconstruction efficiency of 0.12. For reference, we collect the value of spin correlation coefficients with this kinematical selection in Table~\ref{tab:CE}. For the threshold analysis the beamline basis is equivalent to the helicity basis, because the relevant observable, $D={\rm Tr}\ C$ is the same. For the boosted analysis, where the relevant observable is $D_3$, the helicity basis is superior.

\begin{table}[htb]
\begin{center}
\begin{tabular}{ccccccc}
          & $\mtt$     & $|\cosCM|$ & $\mybeta$  & $\sigma$ & $N$ \\
          \hline
Threshold $\nobeta$ & $\leq 390$ & --         & --         & 3.59 pb  & 108000 \\
Threshold $\beta$ & $\leq 390$ & --         & $\leq 0.9$ & 2.76 pb  & 83000 \\
Boosted     & $\geq 800$ & $\leq 0.6$ & --         & 310 fb   & 9400
\\
\hline
\end{tabular}
\caption{Kinematical cuts on $\mtt$ (in GeV), $\cosCM$ and $\mybeta$ used to optimise the figure of merit $S_E$. The fourth column gives the tree-level cross section with the corresponding cuts, and the fifth column the expected number of events after reconstruction (see the text for details).}
\label{tab:cutE}
\end{center}
\end{table}
\begin{table}[htb]
\begin{center}
\begin{tabular}{cccccccc}
            & $C_{kk}$ & $C_{rr}$ & $C_{nn}$ & $C_{kr}$ & $C_{xx}$ & $C_{zz}$ \\
\hline            
Threshold $\nobeta$  & $-0.619$ & $-0.372$ & $-0.568$ & $0.080$  & $-0.606$ & $-0.356$ \\
Threshold $\beta$ & $-0.668$ & $-0.417$ & $-0.593$ & $0.071$  & $-0.632$ & $-0.415$ \\
Boosted     & $0.51$   & $0.64$   & $-0.52$  & $0.15$   & $-0.05$  & $0.76$
\\
\hline    
\end{tabular}
\caption{Parton-level values of the spin correlation coefficients in the helicity and beamline bases with the kinematical selection in Table~\ref{tab:cutE}. In the helicity basis $C_{kr}=C_{rk}$, and in the beamline basis $C_{xx}=C_{yy}$. The rest of coefficients are below 0.01. }
\label{tab:CE}
\end{center}
\end{table}

For each set of cuts we estimate the statistical uncertainty in the measurement of $E$ by performing $n=200$ pseudo-experiments. In each pseudo-experiment we select a random set of $N$ $t \bar t$ events (with $N$ in the last column of Table~\ref{tab:cutE})
and calculate $E$, either from the individual measurements of $C_{ii}$, or with a direct measurement using $D$ or $D_3$.\footnote{The pools of events from which the random samples are selected contain five times more events, therefore the random sets contain some common events. By using different values of $n$ we have checked that the overlap does not bias the determination of the statistical uncertainty. A further check for the uncertainties in the `individual' row is the comparison with those obtained by simple error propagation.}
The mean and standard deviation of the pseudo-experiments are presented in Table~\ref{tab:Esumm}. The standard deviation obtained from the pseudo-experiments is a good estimation of the statistical uncertainty that would be present in such dataset. For the `individual' measurements the uncertainties can also be estimated by simple error propagation. Because the statistical uncertainty in asymmetries is $1/\sqrt{N}$ (provided the asymmetries are small, as is our case), the $C$ coefficients have an uncertainty of $4/\sqrt{N}$, see (\ref{ec:Akl}), and summing in quadrature their uncertainties results in $4 \sqrt{3/N}$. For $N = 108000$, $83000$, $9400$ in Table~\ref{tab:cutE}, this yields $\pm 0.021$, $\pm 0.024$ and $\pm 0.071$, in quite good agreement with the uncertainties in the first line of Table~\ref{tab:Esumm}.

\begin{table}[htb]
\begin{center}
\begin{tabular}{ccccc}
           & Threshold $\nobeta$        & Threshold $\beta$       &   Boosted \\
\hline   
Individual & $0.560 \pm 0.020$ & $0.680 \pm 0.022$ & $0.671 \pm 0.069$ 
\\
Direct     & $0.559 \pm 0.017$ & $0.678 \pm 0.019$ & $0.663 \pm 0.056$ 
\\
\hline   
\end{tabular}
\caption{Values of the entanglement indicator $E$ in (\ref{ec:E1}) obtained from 200 pseudo-experiments with $L = 139~\text{fb}^{-1}$ and the kinematical cuts in Table~\ref{tab:cutE}. The row labeled as `individual' presents results from individual determinations of $C_{kk}$, $C_{rr}$ and $C_{nn}$. The row labeled as `direct' corresponds to results obtained measuring either $D$ (at threshold) or $D_3$ (in the boosted regime).}
\label{tab:Esumm}
\end{center}
\end{table}

At threshold, the use of $\beta$ and the direct determination improve the statistical significance by a factor of 1.27, from $E = 0.560 \pm 0.020$ to $E=0.678 \pm 0.019$. This is quite remarkable, because it would amount to an increase in luminosity by a factor of 1.6. However, for this measurement the statistics are already large, and it is very likely that the uncertainty will be dominated by systematics, so this improvement may not have a great impact in the total uncertainty. In the boosted regime there is an improvement by a factor of 1.23, equivalent to an increase in luminosity by a factor of 1.5, and brings statistical uncertainties below the 10\% level. Provided systematic uncertainties are at the same level, the $5\sigma$ observation of $t \bar t$ entanglement in the boosted regime seems quite feasible. 
In summary, from Table~\ref{tab:Esumm} the two ingredients that can improve the observability of the entanglement are manifest:
\begin{itemize}
\item The application of an upper cut on $\beta$ (only for the threshold analysis), which reduces the $q \bar q$ fraction, enhancing the entanglement and thus the central value of $E$.
\item The direct measurement of $E$, using either the $D$ observable (at threshold) or $D_3$ (in the boosted regime), instead of individually measuring $C_{kk}$, $C_{rr}$ and $C_{nn}$. This direct measurement reduces the statistical uncertainty.
\end{itemize}

\subsection{Observation of CHSH violation}
\label{sec:4.2}

\begin{figure}[htb]
\begin{center}
\includegraphics[width=5.5cm,clip=]{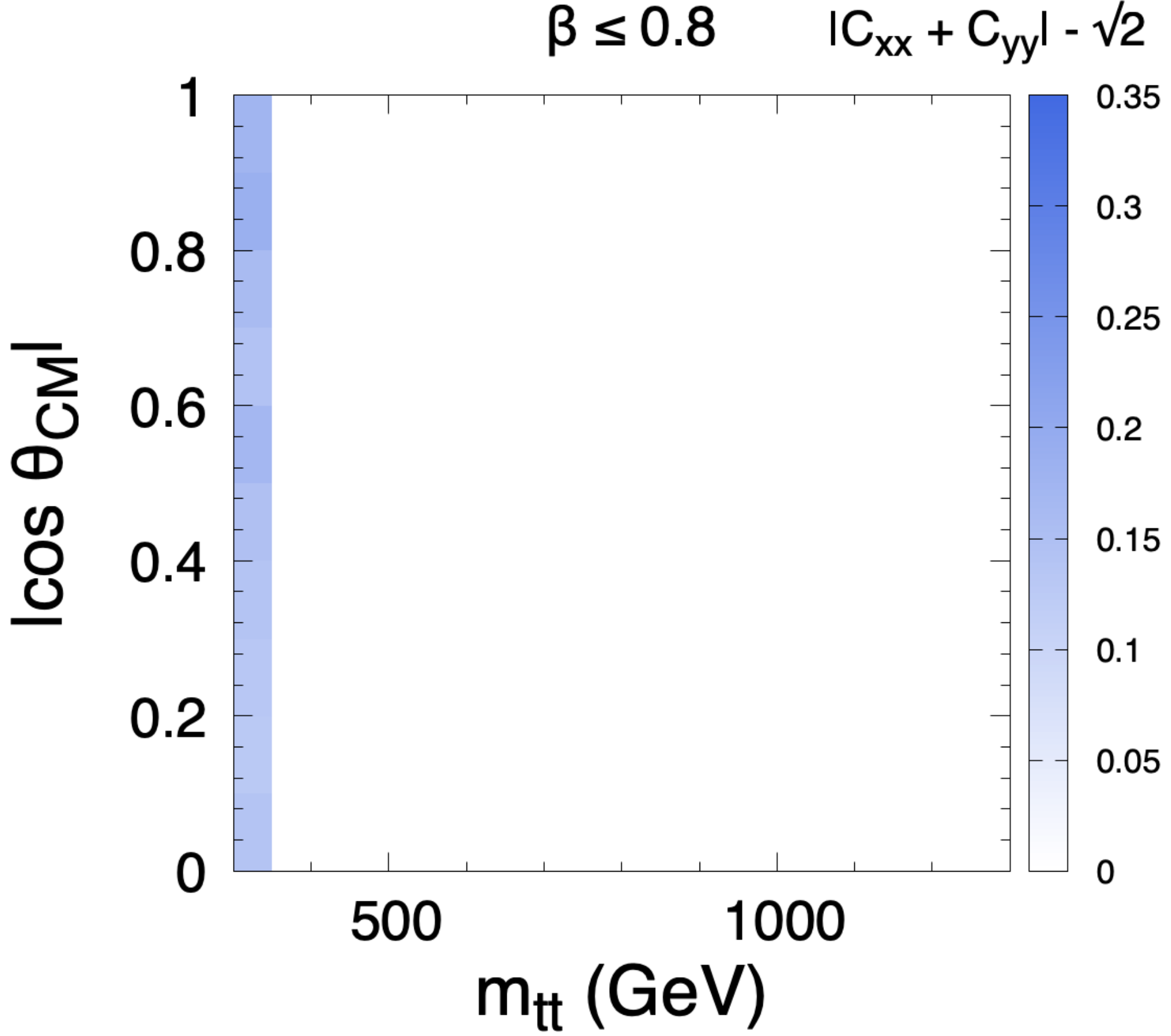}
\caption{Dependence of the $\cancel{\rm CHSH}$ indicator $B_1$ in (\ref{ec:B1}) on $\mtt$ and $\thCM$.}
\label{fig:B1}
\end{center}
\end{figure}

Near threshold, the beamline basis is slightly better to observe a violation of the CHSH inequalities, using as indicator the quantity
\begin{equation}
B_1 \equiv |C_{xx} + C_{yy}| - \sqrt{2} > 0 \,.
\label{ec:B1}
\end{equation}
(For results in the helicity basis see Appendix~\ref{sec:a}.)
Its dependence on $\mtt$ and $\cosCM$ is shown in Fig.~\ref{fig:B1}, requiring $\beta \leq 0.8$.\footnote{In order to reduce statistical fluctuations, in all plots of $\cancel{\rm CHSH}$ indicators these are determined from asymmetries, and we consider $|\cos \theta_\text{CM}|$, in the range $[0,1]$.}
This quantity can be determined by a single measurement (given the fact that $C_{xx} = C_{yy}$ by the symmetry around the beam axis) as proposed in Ref.~\cite{Afik:2022kwm}. However, as we will see in the following, such determination suffers from large statistical uncertainties. The direct determination of the sum $C_{xx}+C_{yy}$ by using an azimuthal asymmetry of the type $A_+$ in (\ref{ec:Aplus}) is much more precise.

\begin{figure}[htb]
\begin{center}
\begin{tabular}{cc}
\includegraphics[width=5.5cm,clip=]{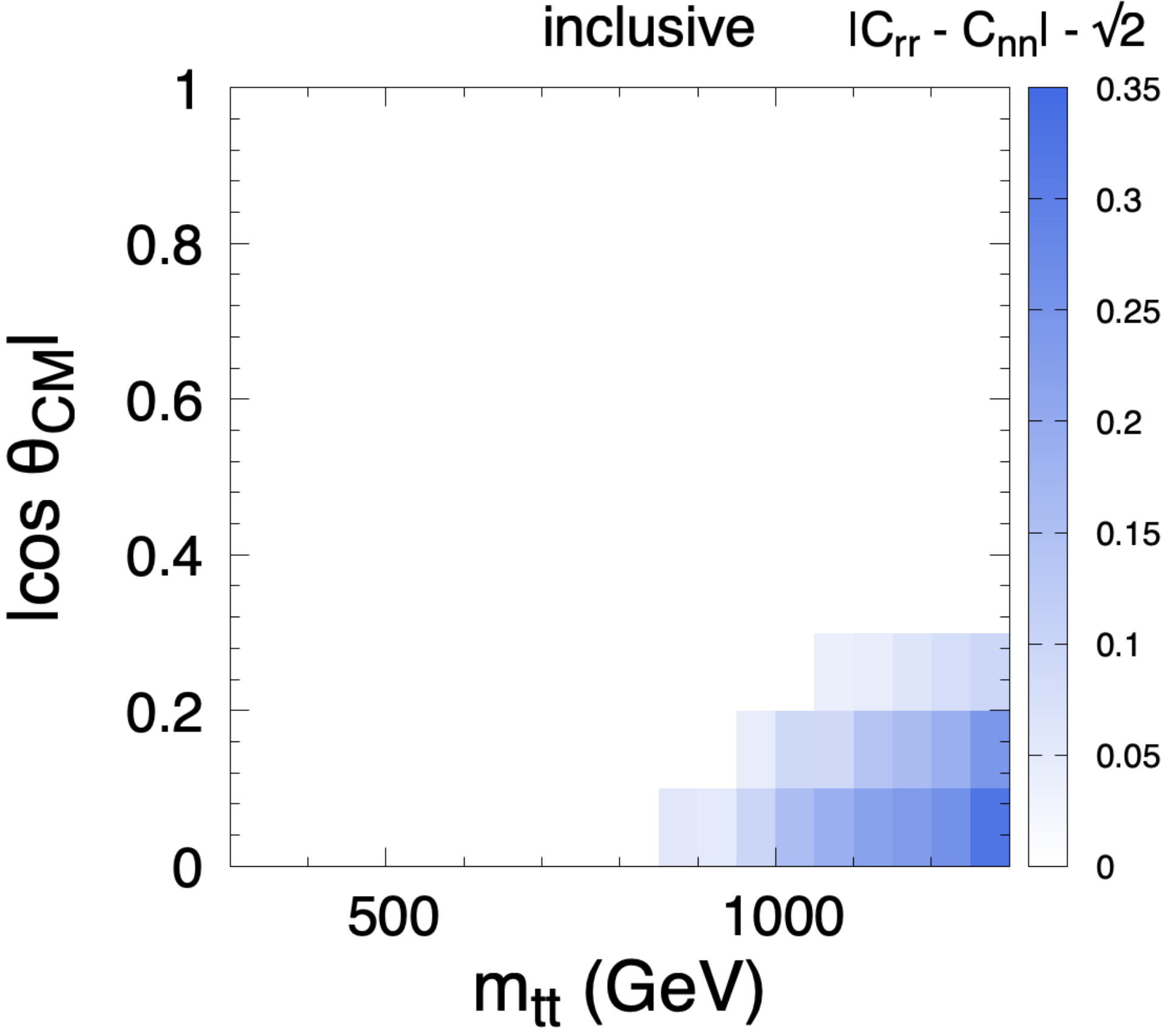} &
\includegraphics[width=5.5cm,clip=]{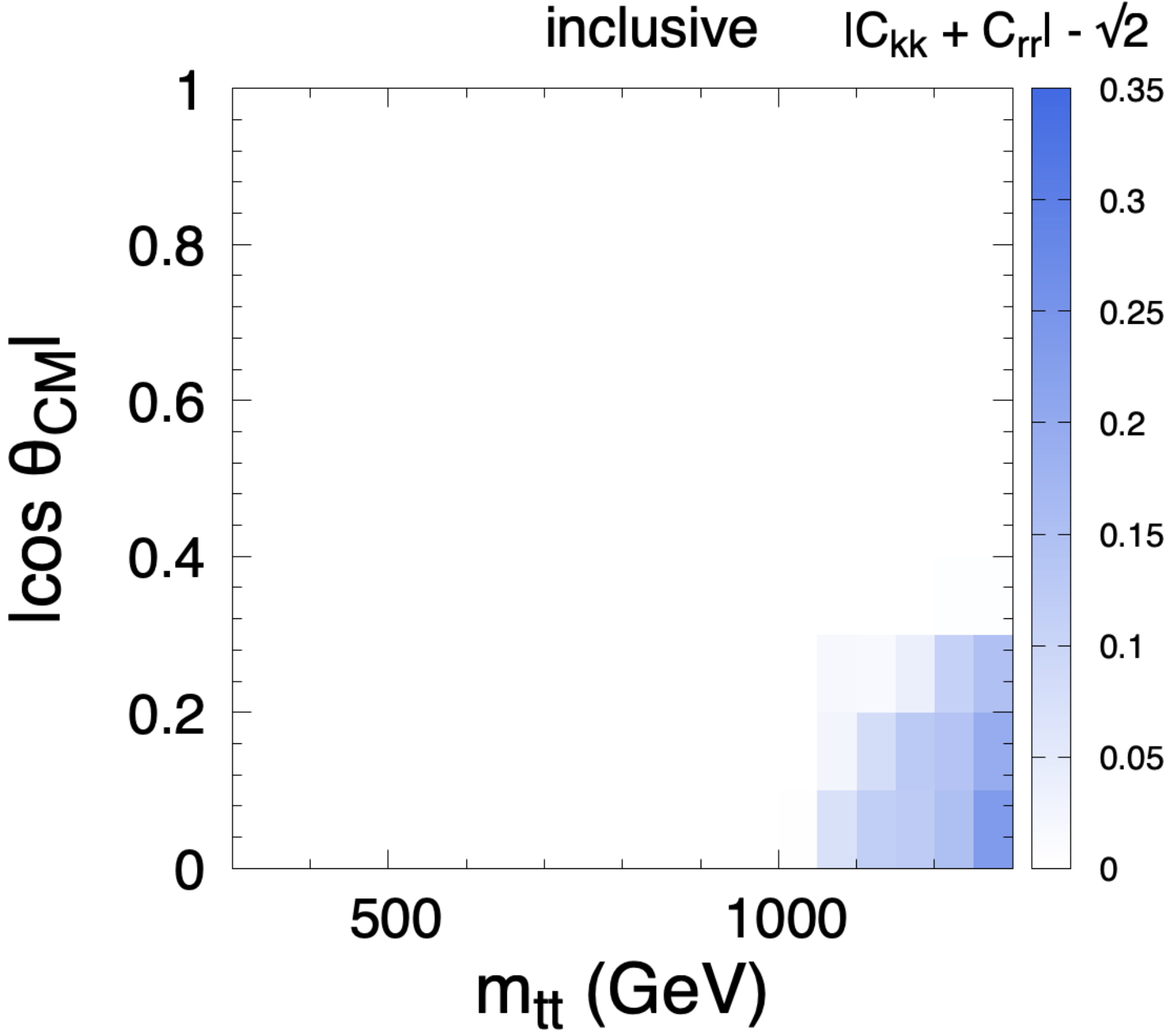} 
\end{tabular}
\caption{Dependence of the $\cancel{\rm CHSH}$ indicators $B_{2,3}$ in (\ref{ec:B23}) on $\mtt$ and $\thCM$.}
\label{fig:B23}
\end{center}
\end{figure}

In the boosted regime the helicity basis is more convenient, and the most useful indicators, by order of importance, are the quantities
\begin{align}
& B_2 \equiv |C_{rr}-C_{nn}| - \sqrt{2} > 0 \,, \notag \\
& B_3 \equiv |C_{kk} + C_{rr}| - \sqrt{2} > 0 \,,
\label{ec:B23}
\end{align}
since $C_{rr}$ and $C_{nn}$ are the two correlation coefficients with largest modulus and they have opposite sign (see Fig.~\ref{fig:C}). Their dependence on $\mtt$ and $\cosCM$ is presented in Fig.~\ref{fig:B23}. We restrict our analysis to $B_2$.
The direct determination of $C_{rr}-C_{nn}$ is obtained from an azimuthal asymmetry of the type $A_-$. 

We use the figure of merit 
\bea
S_B = B_{1,2} \times \sqrt{\sigma}
\eea
to devise a simple set of kinematical cuts, either using $\beta$ or not in the case of the threshold analysis, in order to optimise the statistical significance of $B_{1,2}$. The sets of cuts are collected in Table~\ref{tab:cutB}. 
For each set of cuts we give the tree-level cross section, as well as the number of expected events with a luminosity of 300 fb$^{-1}$, assuming a $K$ factor of 1.8 to normalise the total cross section to next-to-next-to-leading order~\cite{Czakon:2011xx} and using a selection and reconstruction efficiency of 0.12. The value of spin correlation coefficients with these kinematical cuts are presented in Table~\ref{tab:CB}. Clearly, for the threshold analysis the beamline basis is better, while the opposite happens for the boosted analysis. Note also that the off-diagonal spin correlation coefficients are quite small; therefore, the estimators $B_{2,3}$ are very close to the optimal ones.%
\begin{table}[htb]
\begin{center}
\begin{tabular}{ccccccc}
            & $\mtt$      & $|\cosCM|$ & $\mybeta$  & $\sigma$ & $N$ \\
\hline   
Threshold $\nobeta$ & $\leq 353$  & --         & --         & 303 fb   & 19600 \\
Threshold $\beta$ & $\leq 353$  & --         & $\leq 0.8$ & 181 fb   & 11700 \\
Boosted     & $\geq 1000$ & $\leq 0.2$ & --         & 23.3 fb  & 1500 \\
\hline   
\end{tabular}
\caption{Kinematical cuts on $\mtt$ (in GeV), $\cosCM$ and $\mybeta$ used to optimise the figure of merit $S_B$. The fourth column gives the tree-level cross section with the corresponding cuts, and the fifth column the expected number of events after reconstruction (see the text for details).}
\label{tab:cutB}
\end{center}
\end{table}
\begin{table}[htb]
\begin{center}
\begin{tabular}{cccccccc}
            & $C_{kk}$ & $C_{rr}$ & $C_{nn}$ & $C_{kr}$ & $C_{xx}$ & $C_{zz}$ \\
\hline   
Threshold $\nobeta$ & $-0.677$ & $-0.562$ & $-0.712$ & $0.067$  & $-0.719$ & $-0.506$ \\
Threshold $\beta$ & $-0.743$ & $-0.640$ & $-0.761$ & $0.052$  & $-0.767$ & $-0.602$ \\
Boosted     & $0.659$  & $0.874$  & $-0.760$ & $0.037$  & $-0.043$ & $0.878$
\\
\hline   
\end{tabular}
\caption{Parton-level values of the spin correlation coefficients in the helicity and beamline bases with the kinematical selection in Table~\ref{tab:cutB}. In the helicity basis $C_{kr}=C_{rk}$, and in the beamline basis $C_{xx}=C_{yy}$. The rest of coefficients are below 0.01.}
\label{tab:CB}
\end{center}
\end{table}
For each set of cuts we estimate the statistical uncertainty in the measurement of $B_{1,2}$ by performing $n=1000$ pseudo-experiments, and present the results for luminosities of 300 and 3000 fb$^{-1}$ in Tables~\ref{tab:B300} and \ref{tab:B3000}, respectively. The rows  labeled as `individual' give the results obtained from the measurement of individual correlation coefficients (only one of them in the threshold analysis, as proposed in Ref.~\cite{Afik:2022kwm}), while the rows labeled as `direct' give the results from the measurement of azimuthal asymmetries.
From error propagation, one estimates for the `individual' measurements an statistical uncertainty of $4 \sqrt{2/N}$, which yields $\pm 0.040$, $\pm 0.052$ and $\pm 0.15$, respectively, for $N=19600$, $11700$ and $1500$ in Table~\ref{tab:cutB}. The uncertainties quoted in Table~\ref{tab:B300} are slightly larger, but consistent with these estimations.

\begin{table}[htb]
\begin{center}
\begin{tabular}{ccccc}
           & Threshold $\nobeta$       & Threshold $\beta$       & Boosted      \\
\hline   
Individual & $0.021 \pm 0.053$ & $0.119 \pm 0.074$ & $0.218 \pm 0.141$ & \\
Direct     & $0.027 \pm 0.035$ & $0.121 \pm 0.045$ & $0.208 \pm 0.125$ & 
\\
\hline   
\end{tabular}
\caption{Values of the $\cancel{\rm CHSH}$ indicators $B_{1,2}$ in (\ref{ec:B1}), (\ref{ec:B23}) obtained from 1000 pseudo-experiments with with $L = 300~\text{fb}^{-1}$ and the kinematical cuts in Table \ref{tab:cutB}. The row labeled as `individual' gives the results from individual determinations of spin correlation coefficients. The row labeled as `direct' corresponds to results obtained measuring azimuthal asymmetries (see the text for details).}
\label{tab:B300}
\end{center}
\end{table}

\begin{table}[htb]
\begin{center}
\begin{tabular}{ccccc}
           & Threshold $\nobeta$       & Threshold $\beta$       & Boosted      \\
\hline   
Individual & $0.024 \pm 0.017$ & $0.120 \pm 0.021$ & $0.218 \pm 0.041$ \\
Direct     & $0.027 \pm 0.010$ & $0.124 \pm 0.013$ & $0.210 \pm 0.036$ 
\\
\hline   
\end{tabular}
\caption{The same as Table~\ref{tab:B300}, for $L = 3000~\text{fb}^{-1}$.}
\label{tab:B3000}
\end{center}
\end{table}

Near threshold, we observe a great improvement of the statistical significance, e.g. from $0.4\sigma$ to $2.7\sigma$ with 300 fb$^{-1}$, by the combination of a $\beta$ cut and the dedicated observable $A_+$. With 3000 fb$^{-1}$, the statistical uncertainty of $B_1$ is at the 10\% level. Therefore, provided the systematic uncertainties are at the same level, a $5\sigma$ observation of the violation of the CHSH inequalities seems feasible. At the boosted regime the statistical uncertainty in $B_2$ is 17\% for 3000 fb$^{-1}$, thereby allowing for a $5\sigma$ observation of $B_2 > 0$.

\section{Discussion}\label{sec:discussion}

In this work we have investigated novel approaches to improve the observability of entanglement and CHSH violation in top pair production at the LHC. 
The first one is the increase of the $gg$ fraction by a simple upper cut on the $t \bar t$ velocity $\mybeta$, which enhances the entanglement near threshold. The second one is the use of dedicated observables that directly extract the relevant combinations of spin correlation coefficients $C_{ij}$ from data, thereby reducing the statistical uncertainty in the measurements. 

For simplicity, we have not attempted the optimisation of the kinematical cuts in the $(\mtt,\cosCM)$ plane, nor in the $(\mtt,\cosCM,\beta)$ volume, to achieve the highest possible statistical significance in each case. Instead, we have applied simple rectangular cuts on these variables. Further optimisations of the event selection are quite possible, but out of the scope of this work.

As we restrict ourselves to considering statistical uncertainties, our study is performed at the parton level, without detector simulation nor unfolding from detector-level quantities. In particular, the unfolding has been shown to be accurate enough to obtain precise measurements of spin correlation observables~\cite{ATLAS:2016bac,CMS:2019nrx} and template methods similar to the one in Ref.~\cite{Aguilar-Saavedra:2021ngj} could also be exploited.
Backgrounds in the dilepton channel have also been shown to be small~\cite{ATLAS:2016bac,CMS:2019nrx}, and have been ignored in our analysis. We have used a fixed CM energy of 13 TeV, even if in Run 3 of the LHC it will be raised to 13.6 TeV, thereby providing slightly larger samples than the ones assumed here.

Statistical uncertainties are mainly driven by the number of events obtained after suitable kinematical cuts. We have used an overall efficiency for a typical event selection and $t \bar t$ reconstruction consistent with the one found in other works~\cite{Severi:2021cnj}. Possible departures from this reference value would modify the size of all our event samples by the same amount. As the goal of our work is to investigate the improvement brought by the use of $\beta$ and dedicated observables, our conclusions are quite robust even if the selection and reconstruction efficiency has some departure from the reference value used here.

The methods introduced here to enhance the sensitivity are especially effective when statistical uncertainties are important, namely
\begin{itemize}
\item[(i)] for the observation of entanglement in the boosted regime;
\item[(ii)] for the observation of CHSH violation both at threshold and in the boosted regime.
\end{itemize}
In the boosted region, a cut on $\beta$ does not provide any advantage, so the central value (both for entanglement and CHSH violation) remains the same, but the statistical uncertainty decreases thanks to the dedicated observables. 
We find a moderate improvement in the significance for the observation of both phenomena. This strengthens in particular the conclusion of Ref.\cite{Severi:2021cnj} in the sense that entanglement can be measured with Run 2 data.

On the other hand, the greatest improvement occurs for the measurement of the CHSH violation near threshold. With $L = 300~\text{fb}^{-1}$, we find that the observable proposed in Ref.~\cite{Afik:2022kwm}
only allows for $0.4\sigma$ statistical significance, while the asymmetry  introduced here, together with the use of $\beta$, enhances the sensitivity to $2.7\sigma$.

In this work we have not addressed systematic uncertainties ---nor it is possible without a full detector simulation. Although these uncertainties will be important, we believe that our proposal for optimisation will be quite useful for future analyses. The application of an upper cut on $\beta$ selects the events that are more central, where the detector resolution is better, and is not expected to degrade the systematic uncertainty, but on the contrary, it may improve systematics. Concerning the use of dedicated observables, there is no a priori reason why the systematic uncertainties should be larger for them. One would even expect that uncertainties may be smaller. For example, the determination of $C_{kk}+C_{rr}-C_{nn}$ from its individual terms involves a precise reconstruction of the three axes $(\hat k,\hat r,\hat n)$ (in addition to the lepton momenta) while the determination from $D_3$ is only sensitive to the reconstruction of the $(\hat k, \hat r)$ plane. 

In conclusion, the methods outlined here may be quite useful for entanglement measurements in the boosted regime, where already with Run 2 data a measurement seems quite feasible. For CHSH violation more data is required, and $3 \sigma$ evidence would already be feasible with Run 3.

\section*{Acknowledgements}

We thank A. Bernal, J.M. Moreno for very useful discussions, and the authors of Ref.~\cite{Severi:2021cnj} for discussions and information about their work.
This work has been supported by 
 the grants IFT Centro de Excelencia Severo Ochoa 
CEX2020-001007-S,  PID2019-110058GB-C21 and PID2019-110058GB-C22, funded by MCIN/AEI/10.13039/501100011033 and by ERDF, 
and by FCT project CERN/FIS-PAR/0004/2019. 

\appendix
\section{CHSH violation in the helicity basis at threshold}
\label{sec:a}

\begin{figure}[htb]
\begin{center}
\begin{tabular}{cc}
\includegraphics[width=5.5cm,clip=]{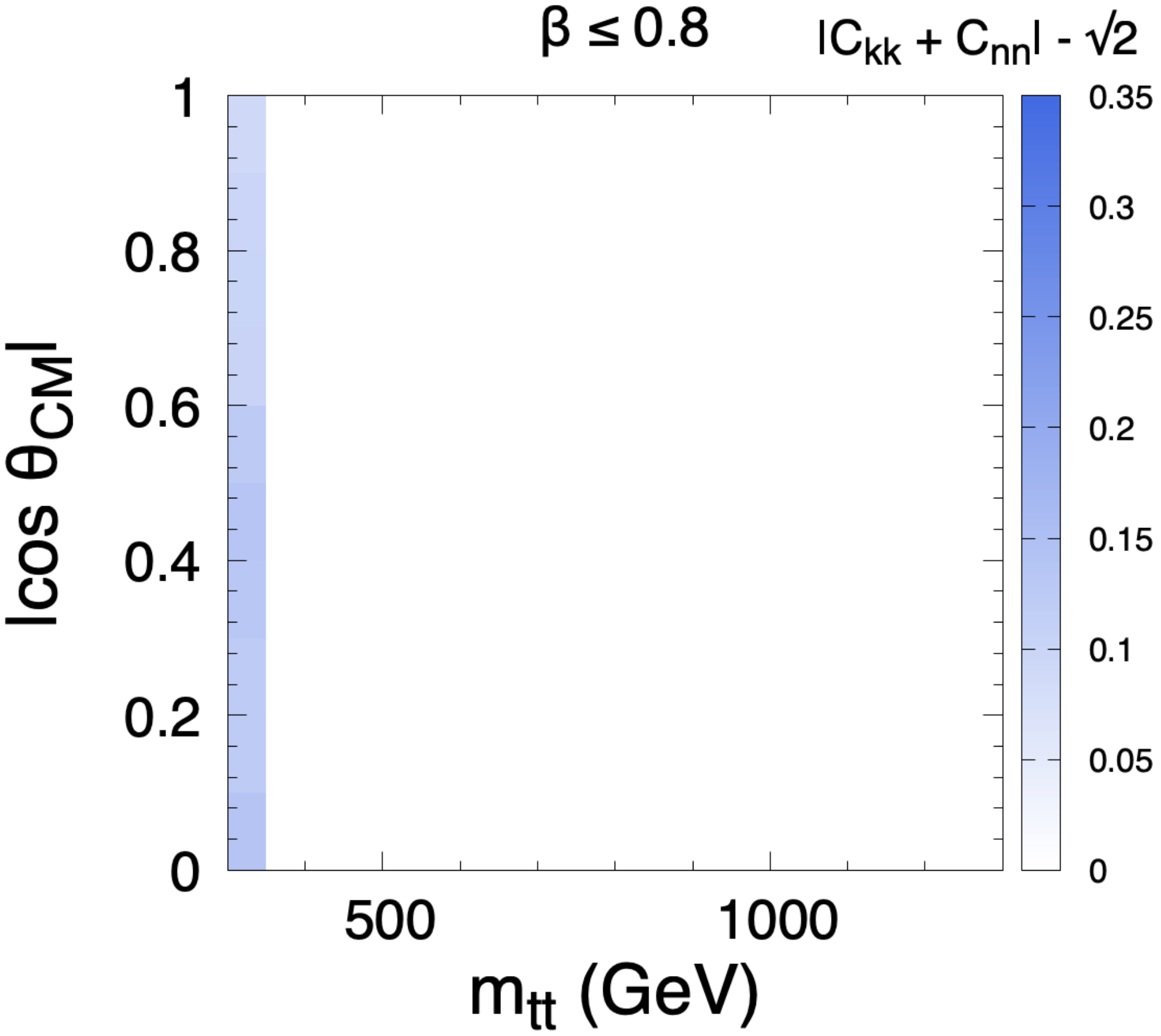} & 
\includegraphics[width=5.5cm,clip=]{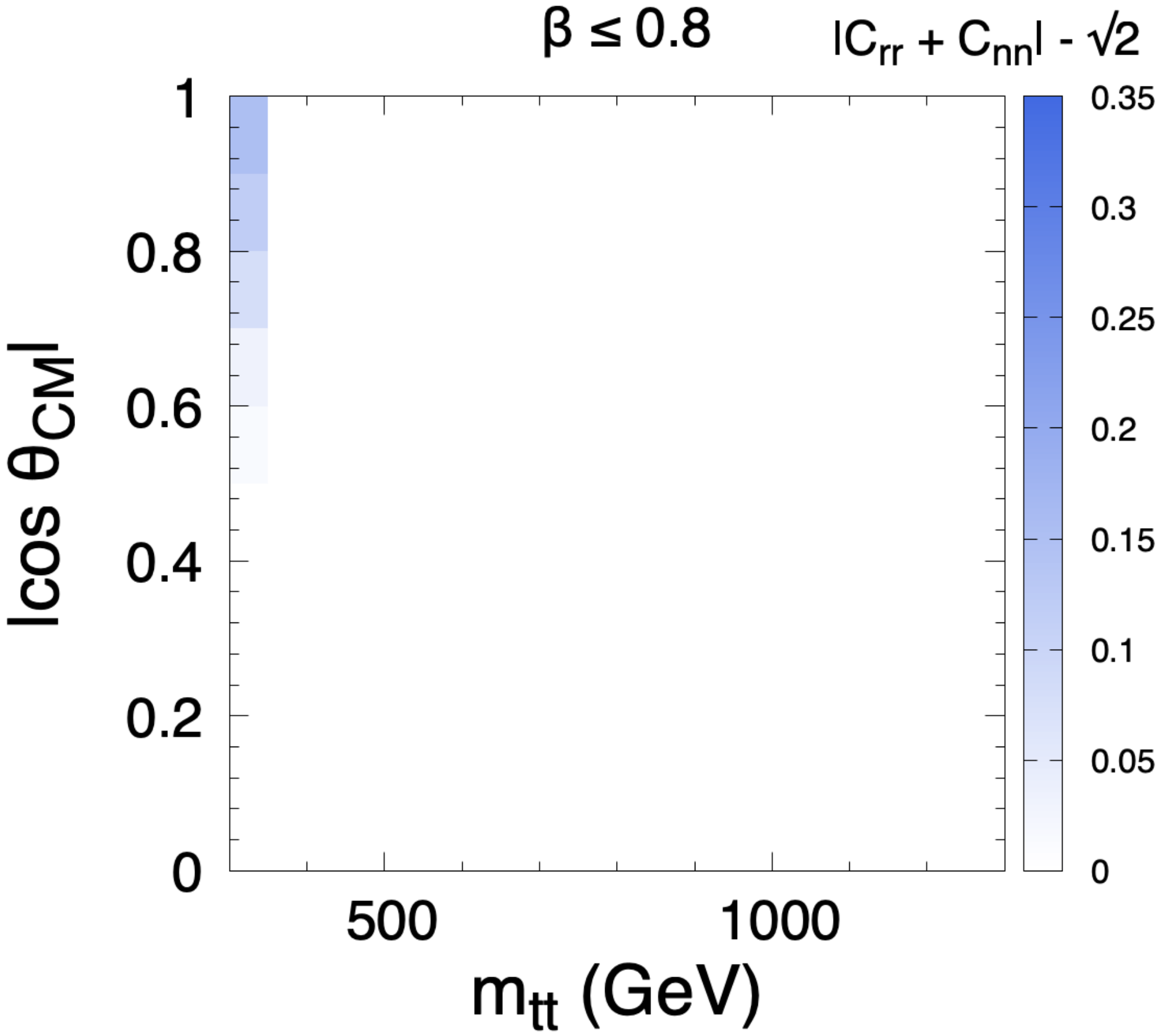} 
\end{tabular}
\caption{Dependence of the indicators $B_{4,5}$ in (\ref{ec:B45}) on $\mtt$ and $\thCM$.}
\label{fig:Bother}
\end{center}
\end{figure}

In the helicity basis the CHSH inequalities are also violated near threshold. The most significant indicators are, by order of importance,
\begin{align}
& B_4 \equiv |C_{kk} + C_{nn}| - \sqrt{2} > 0 \,, \notag \\
& B_5 \equiv |C_{rr} + C_{nn}| - \sqrt{2} > 0 \,, 
\label{ec:B45}
\end{align}
Their dependence on $\mtt$ and $\cosCM$ is shown in Fig.~\ref{fig:Bother}, after applying an upper cut $\mybeta \leq 0.8$. We focus on $B_4$ and follow the same procedure as in Section~\ref{sec:4.2}, requiring $\mtt \leq 355$ GeV. Performing $n=1000$ pseudo-experiments, we find that when using an asymmetry of the type $A_+$, the extracted values are
\begin{align}
& B_4 = 0.085 \pm 0.038 \quad L=300~\text{fb}^{-1} \,, \notag \\
& B_4 = 0.084 \pm 0.012 \quad L=3000~\text{fb}^{-1} \,.
\end{align}
The statistical significance of non-zero $B_4$ is $2.2\sigma$ and $7\sigma$, respectively. For comparison, in the beamline basis (Tables~\ref{tab:B300} and \ref{tab:B3000}) one obtains for $B_1$ significances of $2.7\sigma$ and $9.5\sigma$.

\section{Comparison with previous work}
\label{sec:b}

In this appendix, rather than addressing the improvements of the strategy presented here with respect to previous approaches (which has already be done in the main body of the article and the discussion), we focus on the numerical (dis)agreement found within our setup, when we follow a strategy similar to those works, in particular Ref.~\cite{Severi:2021cnj}. Note that while our results are directly obtained at the parton level from the generator, those in Ref.~\cite{Severi:2021cnj} are obtained after hadronisation and showering by {\scshape Pythia}~\cite{Sjostrand:2007gs}, detector simulation with {\scshape Delphes}~\cite{deFavereau:2013fsa}, reconstruction of the $t \bar t$ kinematics and unfolding to parton level of the measured observables.

In order to compare our results for entangement and CHSH violation 
we implement their kinematical cuts on the $(\mtt,\cosCM)$ plane, namely their selections labeled as `strong'. We use the cross sections reported in Ref.~\cite{Severi:2021cnj} for these kinematical selections, given in the first column of Table~\ref{tab:comp}, which are similar to the ones we obtain after the multiplication by the $K$ factor. Taking the event selection and reconstruction efficiencies in the second column\footnote{We thank the authors of  Ref.~\cite{Severi:2021cnj}  for providing these numbers.}, the number of events for luminosities $L=139~\text{fb}^{-1}$ (entanglement) and $L=350~\text{fb}^{-1}$ (CHSH violation) are given in the third column.

For the entanglement (CHSH violation) analyses we perform $n=200$ ($n=1000$)
 pseudo-experiments and give in the fifth column of Table~\ref{tab:comp} the central value and standard deviation of $E+1=|C_{kk}+C_{rr}|-C_{nn}$ and $2 + \sqrt{2} B_2 = \sqrt{2}|C_{rr}-C_{nn}|$. In all cases we determine these quantities from the individual spin correlation coefficients, and subsequently perform the corresponding sums. The results from Ref.~\cite{Severi:2021cnj} are given in the last column, where the uncertainties are statistical but also include some systematic effects arising from the bin migrations caused by imperfect reconstruction. Given these differences, we find that the uncertainties found in the two analyses are in good agreement.

\begin{table}[htb]
\begin{center}
\begin{small}
\begin{tabular}{rcccccc}
& $\sigma$ & eff & $N$ & indicator & this work & Ref.~\cite{Severi:2021cnj} \\
\hline   
Entanglement threshold & 10 pb  & 0.08 & 97300 & $|C_{kk}+C_{rr}|-C_{nn}$  & $1.476 \pm 0.020$ & $1.38 \pm 0.02$ \\
Entanglement boosted   & 0.9 pb & 0.075 & 9400  & $|C_{kk}+C_{rr}|-C_{nn}$  & $1.574 \pm 0.065$ & $1.42 \pm 0.10$ \\
CHSH boosted           & 60 fb  & 0.011 & 231  & $\sqrt{2}|C_{rr}-C_{nn}|$ & $2.277 \pm 0.52$  & $2.30 \pm 0.76$
\\
\hline   
\end{tabular}
\end{small}
\caption{Comparison of entanglement and CHSH violation indicators with Ref.~\cite{Severi:2021cnj}.}
\label{tab:comp}
\end{center}
\end{table}


In order to assess the differences in the central values we give in Table~\ref{tab:comp2} the parton-level values of the entanglement indicators using the Monte Carlo settings described in Section~\ref{sec:4} (second column) as well as with samples generated using the default PDFs (NNPDF 2.3) and factorisation and renormalisation scales in {\scshape MadGraph} (third column). The values are quite consistent, with a small decrease in $|C_{kk}+C_{rr}|-C_{nn}$ at threshold, which may explain the difference with respect to Ref.~\cite{Severi:2021cnj}. For $|C_{kk}+C_{rr}|-C_{nn}$ in the boosted regime, the discrepancy in the central values appears to have a statistical origin.

\begin{table}[htb]
\begin{center}
\begin{tabular}{rccccc}
& Indicator & This work & Default \\
\hline   
Entanglement threshold & $|C_{kk}+C_{rr}|-C_{nn}$  & $1.470 \pm 0.004$ & $1.426 \pm 0.004$ \\
Entanglement boosted   & $|C_{kk}+C_{rr}|-C_{nn}$  & $1.576 \pm 0.013$ & $1.588 \pm 0.013$ \\
CHSH boosted           & $\sqrt{2}|C_{rr}-C_{nn}|$ & $2.264 \pm 0.009$ & $2.244 \pm 0.010$
\\
\hline   
\end{tabular}
\caption{Comparison of entanglement and CHSH violation indicators at the parton level, with two Monte Carlo settings. The uncertainties are from Monte Carlo statistics.}
\label{tab:comp2}
\end{center}
\end{table}
Let us finally mention that the simulation in Ref.~\cite{Severi:2021cnj} includes all Feynman diagrams (around 5000) contributing to the $\ell^+ \nu b \ell^- \nu \bar b$ final state, even those without intermediate top quarks --- whose contribution is mentioned there to be small. In this regard, we believe that, if CHSH inequalities are to be tested for (hypothetical) spin measurements on entangled $t \bar t$ pairs, only the contributions that actually have top quarks must be included.
A conceptually different issue is whether backgrounds to this process, actually without top quarks, can dilute or even {\em fake} a violation of the CHSH inequalities for top quarks. Anyway, as already mentioned, those backgrounds have been shown in Ref.~\cite{Severi:2021cnj} to be small after the $t \bar t$ reconstruction.

\end{document}